\documentclass[]{aa}    
\usepackage{psfig}

\def\secondip{\hbox{\rlap{\hbox{.}}\hbox{$''$}}}

\def\gradip{\hbox{\rlap{\hbox{.}}\raise 5.truept \hbox{{\small $\circ$}}}}
\def\msun{\hbox{$M_\odot$}}
\def\und{\underline}
\def\msun{\hbox{$M_\odot$}}
\catcode`\@=11
\def\gsim{\ifmmode{\mathrel{\mathpalette\@versim>}}
    \else{$\mathrel{\mathpalette\@versim>}$}\fi}
\def\lsim{\ifmmode{\mathrel{\mathpalette\@versim<}}
    \else{$\mathrel{\mathpalette\@versim<}$}\fi}
\def\@versim#1#2{\lower 2.9truept \vbox{\baselineskip 0pt \lineskip
    0.5truept \ialign{$\m@th#1\hfil##\hfil$\crcr#2\crcr\sim\crcr}}}
\catcode`\@=12

\begin{document}

\title{Age and Metallicity Distribution of the Galactic Bulge from
Extensive Optical and Near-IR Stellar Photometry
\thanks{Based on observations collected at the European Southern
Observatory, La Silla, Chile, obtained from the ESO/ST-ECF Science
Archive Facility, and on observations with the NASA/ESA Hubble
Space Telescope, obtained at the Space Telescope Science Institute,
operated by AURA Inc. under contract to NASA}}

\author{M. Zoccali\inst{1}, A. Renzini\inst{1}, S. Ortolani\inst{2},
L. Greggio\inst{3}, I. Saviane\inst{4}, S. Cassisi\inst{5}, M. Rejkuba\inst{1},
B. Barbuy\inst{6}, R.M. Rich\inst{7}, E. Bica\inst{8}}

\institute{
European Southern Observatory, Karl Schwarzschild Strasse 2, D-85748,
Garching bei M\"unchen, Germany;  mzoccali@eso.org, arenzini@eso.org,
mrejkuba@eso.org
\and
Universit\`a di Padova, Dipart. di Astronomia, Vicolo dell'Osservatorio 2,
I-35122 Padova, Italy; ortolani@pd.astro.it
\and
INAF, Osservatorio Astronomico di Padova, Padova, Italy; greggio@pd.astro.it
\and
European Southern Observatory, Alonso de Cordova 3107, Santiago 19, Chile;
isaviane@eso.org
\and
INAF, Osservatorio Astronomico Collurania, I-64100, Teramo, Italy;
cassisi@te.astro.it
\and
Universidade de S\~ao Paulo, CP 3386, S\~ao Paulo 01060-970, Brazil;
barbuy@astro.iag.usp.br
\and
Department of Physics and Astronomy, Division of Astronomy and Astrophysics,
University of California, Los Angeles, CA 90095-1562; rmr@astro.ucla.edu
\and
Universidade Federal do Rio Grande do Sul, Dept. de Astronomia, CP 15051,
Porto Alegre 91501-970, Brazil;
bica@if.ufrgs.br
}

\date{Received date / Accepted date}

\authorrunning{M. Zoccali et al.}

\titlerunning{Age and Metallicity Distribution of the Galactic bulge}

\abstract{
We present a new determination of the metallicity distribution, age,
and luminosity function of the Galactic bulge stellar population.  By
combining near-IR data from the 2MASS survey, from the SOFI imager at
ESO NTT and the NICMOS camera on board HST we were able to construct
color-magnitude diagrams (CMD) and luminosity functions (LF) with
large statistics and small photometric errors from the Asymptotic
Giant Branch (AGB) and Red Giant Branch (RGB) tip down to $\sim 0.15
M_\odot$. This is the most extended and complete LF so far obtained
for the galactic bulge.  Similar near-IR data for a disk control field
were used to decontaminate the bulge CMDs from foreground disk stars,
and hence to set stronger constraint on the bulge age, which we found
to be as large as that of Galactic globular clusters, or $\gsim 10$
Gyr. No trace is found for any younger stellar population. Synthetic
CMDs have been constructed to simulate the effect of photometric
errors, blending, differential reddening, metallicity dispersion and
depth effect in the comparison with the observational data. By
combining the near-IR data with optical ones, from the Wide Field
Imager at the ESO/MPG 2.2m telescope, a disk-decontaminated
$(M_K,V-K)$ CMD has been constructed and used to derive the bulge
metallicity distribution, by comparison with empirical RGB
templates. The bulge metallicity is found to peak at near solar value,
with a sharp cutoff just above solar, and a tail towards lower
metallicity that does not appreciably extend below [M/H]$\sim-1.5$.
}

\maketitle

\section{Introduction}

Galactic spheroids, i.e., elliptical galaxies and the bulges of spiral
galaxies, contain a large fraction,
probably the majority, of all the stellar mass in the local universe (Persic
\& Salucci 1992; Fukugita, Hogan, \& Peebles 1998).  Understanding
their formation and evolution is therefore crucial to understand
galaxy formation in general, and to reconstruct the whole cosmic star
formation history.  However, it is only in the bulge of our own Galaxy
that we can resolve the stellar population all the way to the bottom
of the main sequence (MS), construct accurate color-magnitude diagrams
(CMD) and luminosity functions (LF), and then address a series of
issues of great importance for a better understanding of galactic
spheroids in general. Such issues include the direct measure of some
of the fundamental  ingredients of galaxy models, like the initial
mass function (IMF), the mass-to-light ratio and quantities
such as the stellar age and metallicity distributions (MD).

Efforts in these directions have been quite numerous in recent years,
and reviewing all of them is well beyond the scope of this
paper. However, here are briefly mentioned some of the most recent
attempts at determining the age, IMF and MD of the bulge stellar
population.

The only published study of the bulge MD based on high resolution
(R=17,000) spectra remains that by McWilliam \& Rich (1994; hereafter
M\&R). They analyzed a sample of 11 red giants in Baade's Window, and
used these data to re-calibrate a previous bulge MD based on low
resolution spectra for 88 stars (Rich 1988). The resulting MD is centered
around [Fe/H]=$-0.2$, with some 34$\%$ of the stars above solar
metallicity, and no stars below [Fe/H]$\simeq -1.3$.

The bulge MD has also been derived using lower resolution spectra.
Sadler et al. (1996) measured line strength $<\!{\rm Fe}\!>$ indices in
low resolution ($R\sim1000$) spectra of a sample of 322 bulge
giants. These indices were calibrated vs.\ [Fe/H] following Faber et
al. (1985). Although the derived [Fe/H] abundances for the few stars
in common with M\&R are in good agreement, the global MD by Sadler et
al. (1996) is symmetric around [Fe/H]=0, and shows a population of
stars with metallicity above solar more prominent than in any other
study, which may result from having applied the $<\!{\rm Fe}\!>$ vs. [Fe/H]
calibration beyond its range of validity. More recently, Ramirez et
al. (2000) derived the bulge MD by measuring the equivalent width of
Ca, Na and CO lines in a sample of $110$ M giants observed with
resolutions from R$\sim 1300$ to $4800$. They converted equivalent
widths into [Fe/H] by means of the Frogel et al. (2001) calibration,
based on Galactic globular clusters (GCs) of independently known
[Fe/H].  This assumes that the $\alpha$-element enhancement of
the bulge is the same as that of the clusters.  The MD obtained in
this case is similar to that of M\&R, but with a sharper peak at
[Fe/H]=0.  The appreciable differences among the MDs derived in  these
studies are likely to arise from uncertainties in the calibration of
the low resolution indices, that may introduce important systematics.

Dating bulge stars is complicated by several factors, such as
crowding, depth effects, variable reddening, metallicity dispersion,
and contamination by foreground disk stars.  From WFPC1 observations
of the bulge field known as Baade's Window, Holzman et al. (1993)
inferred a dominant intermediate age component in the Galactic
bulge. Shortly after the refurbishment of HST, two metal rich globular
clusters of the bulge (NGC 6528 and NGC 6553) were observed with WFPC2
(Ortolani et al. 1995, hereafter Paper I). These clusters are
respectively at $\sim4^\circ$ and $\sim6^\circ$ from the galactic
center, and their overall metallicity [M/H] is about solar (Barbuy et
al. 1999; Cohen et al. 1999), close to the average for stars in
Baade's Window (M\&R). Like most other clusters within $\sim3$ kpc
from the Galactic center, they belong to the population of bulge
globular clusters that have age, kinematical properties and
metallicity distribution that are indistinguishable from those of
bulge stars (e.g., Minniti 1995; Paper I; Cot\'e 1999). Hence, we do
not hesitate to refer to them as {\it bulge globular clusters} (see
also Forbes, Brodie \& Larsen 2001; Harris 1976, 2001; Zinn 1996). In
the case of NGC 6528 and NGC 6553 the bulge membership is also
confirmed by the proper motion of the two clusters, in both cases well
within the proper motion distribution of bulge stars (Zoccali et
al. 2001a; Feltzing, Johnson, \& de Cordova 2002).  In Paper I it was
shown that {\it i}) the two clusters have virtually identical CMDs,
indicating that they have the same age and metallicity; {\it ii})
their magnitude difference between the horizontal branch (HB) {\it
clump} and the MS turnoff is virtually identical to that of the inner
halo globular cluster NGC~104 ([Fe/H]=$-0.7$), and {\it iii}) the LF
of NGC 6528 (the least reddened of the two clusters) is virtually
identical to the LF of bulge stars in Baade's Window, when allowance
is made for the distribution of bulge star distance along the line of
sight.  From all these evidences Ortolani et al. concluded that {\it
i}) the bulge underwent rapid chemical enrichment to solar abundance
and beyond, very early in the evolution of the Milky Way (MW) Galaxy;
{\it ii}) the bulk of bulge stars formed nearly at the same time as
the halo globular clusters, and {\it iii}) no more than $\sim 10\%$ by
number of the bulge population can be represented by intermediate age
stars. These conclusions have been further strengthened after a
statistical decontamination of the bulge CMD from the foreground disk
stars (Feltzing \& Gilmore 2000).

Dating the bulge via the MS turnoff also allows to check the
usefulness of asymptotic giant branch (AGB) stars brighter than the
red giant branch (RGB) tip as age indicators.  Indeed, being such AGB
stars much brighter than the MS turnoff, they have often been used in
the attempt to date other galactic spheroids (such as M32 and the
bulge of M31), sometimes inferring intermediate ages for some of these
systems (e.g, Davidge and Nieto, 1992; Freedman 1992; Elston and
Silva, 1992).  However, the existence of AGB stars brighter than the
RGB tip does not ensure a population to be of intermediate age: the
old, metal rich ([Fe/H]$\gsim-0.7$) globular clusters of the MW bulge
and inner halo (such as, e.g., NGC~6553 and NGC~104) do indeed
contain some AGB stars that are $\sim 1$ magnitude brighter than the
RGB tip (Frogel and Elias 1988; Guarnieri, Renzini, \& Ortolani
1997). In any event, the RGB+AGB luminosity function conveys important
information concerning the stellar population of the bulge, and its
better understanding is important for the interpretation of the
observations of nearby spheroids.

At the opposite extreme of the luminosity range, the LF of the lower
MS is the only part of the LF that depends on the IMF, and therefore
it allows the determination of this important property of stellar
populations.  Recent near-infrared data obtained with HST+NICMOS, have
provided the deepest ever $J,H$ photometry in the bulge, and allowed
the determination of the IMF from $\sim 1 M_\odot$ down to $\sim
0.15M_\odot$ (Zoccali et al. 2000, hereafter Paper II).  The IMF
resulted to be quite {\it flat}, with a slope $-1.33\pm0.07$ (compared
to $-2.35$ for the Salpeter IMF), with a hint of a steepening toward
the high mass end (see also Holzman et al. 1998).

The study of the NICMOS field extended to only 408 square arcsec
\footnote{The size of the NICMOS field was erroneously quoted as 506
square arcsec in Paper~II.}, and therefore the number of stars on the
upper MS is too small to allow any reliable location of the MS
turnoff, while the evolutionary stages beyond the turnoff are not
sampled at all.  Thus, in Paper II an attempt was made to extend the
main sequence LF to the evolved stars, using near-infrared data from
the literature. A {\it complete} LF, extending from the bottom of the
MS all the way to the AGB, was constructed by matching the NICMOS
luminosity function of the lower MS, to the Tiede, Frogel \& Terndrup
(1995) LF extending from slightly above the turnoff to slightly above
the HB, and finally to the Frogel \& Whitford (1987) LF for the bulge
M giants above the HB.  While the resulting composite LF was
constructed with the best data available in each luminosity range,
several limitations were also obvious.  Specifically, the MS turnoff
region was not well sampled by neither the NICMOS data from Paper II
(too few stars in the small field), nor by the Tiede et al. LF (not
deep enough to reach the turnoff). Moreover, the upper RGB and AGB
from Frogel \& Whitford (1987) included only the M stars selected from
an objective prism survey, hence omitting any luminous K giants that
may be present in the explored bulge field. Finally, the NICMOS
observations were conducted in a field $6^\circ$ from the Galactic
center, while the two other datasets were relative to Baade's Window,
at $4^\circ$ from the center.

In this paper we present a thorough attempt at overcoming the current
limitations of the available CMDs and LFs of the bulge, thereby
producing {\it state of the art} CMDs and LFs used for new
determinations of the age and metallicity distribution of the bulge.
This approach is based on new near-IR ($J,H,K_s$) data obtained with
SOFI at the ESO New Technology Telescope (NTT), as well as on optical
($V,I$) data obtained with the Wide Field Imager (WFI) at the ESO/MPG
2.2m telescope. These data are then coupled with the NICMOS data from
Paper II for the lower MS, and with the 2MASS near-IR data for the
upper RGB and AGB (van Dyk, 2000; Carpenter, 2001; and references
therein).

The paper is organized as follows: Observations and data reduction
procedures are presented in Section~2, the CMDs are displayed and
discussed in Section~3, while Section~4 is devoted to the derivation
of the metallicity distribution.  The luminosity function of the bulge
is constructed and discussed in Section~5, while Section~6 is devoted
to determining the age of the bulge, and Section~7 to a discussion of
the RGB tip in the different bands. Finally, the main results are
again summarized in Section~8, along with some inferences on the
formation of the Galactic bulge and spheroid, also in the context of
the current evidence on the formation of galactic spheroids in
general.

\section{Observations and Data Reduction}
\label{sec_obs}

The data presented here come from the combination of near-IR $J,H,K_s$
observations conducted with SOFI@NTT, implemented with public data
from the 2MASS survey and with NICMOS data from Paper II, plus optical
$V,I$ images taken with the WFI@2.2m as part of the EIS PRE-FLAMES
survey (Momany et al. 2001; and references therein), and retrieved
from the ESO archive. The relevant technical information about the
observations are reported in Table~\ref{logobs}, while details on the
observational strategy and data reductions are described here below.

\subsection{Near-IR data}

A bulge mosaic field centered at ($l,b$)=($0.277,-6.167$)
(RA=18:11:13, DEC=$-$31:43:49; J2000) was observed with SOFI@NTT,
through the filters $J,H,K_s$, in the nights of 9--12 June 2001.  This
particular field was selected for the present study because it
includes the field already covered by the NICMOS observations
(Paper~II), which in turn was chosen for having a reddening
($E(B-V)=0.47$) as low as that of the most widely studied Baade's
Window while being less crowded.  Indeed, being $\sim 2^\circ$ further
away from the Galactic center the density of stars is lower by a
factor $\sim 2$, which allows more accurate photometry especially at
very faint magnitudes.

In order to optimize the photometry of both the turnoff region and the
upper RGB, the observations were split in short and long exposures,
each with a different field objective, and therefore mapping areas of
different sizes.  For a good sampling of RGB+AGB stars, short
exposures were obtained for a $8\farcm3\times8\farcm3$ region, with a mosaic
of four fields of the low resolution camera, which has a pixel size of
$0\farcs292$ and field of view of $4\farcm9\times4\farcm9$ (hereafter SOFI-LARGE
field). In order to better sample the PSF, a second mosaic of four
deeper exposures mapping a smaller area ($3\farcm9\times3\farcm9$) was secured
with the same camera coupled with the focal elongator, therefore
yielding a pixel size of $0\farcs144$ and a $2\farcm5\times2\farcm5$ field of view
(hereafter SOFI-SMALL field). An overlap of $\sim 20\%$ was always
present among the four fields of each mosaic.  

\begin{table*}
\begin{center}
\caption[1]{Log of the observations}
\begin{tabular}{cccccccc}
\hline
\noalign{\smallskip}
Date & Object & Camera & RA & DEC & Filter & Exptime \\
\noalign{\smallskip}
\hline
\noalign{\smallskip}
11 Jun 2001 & bulge & SOFI-LARGE & 18:11:05.0 & $-$31:45:49 & J  & 36s   \\[-2pt]
  $''$	      &  $''$ & $''$       &    $''$    &     $''$    & H  & 36s   \\[-2pt]
  $''$	      &  $''$ & $''$       &    $''$    &     $''$    & Ks & 36s   \\[-2pt]
  $''$	      &  $''$ & $''$       & 18:11:22.0 & $-$31:45:49 & J  & 36s   \\[-2pt]
  $''$	      &  $''$ & $''$       &    $''$    &     $''$    & H  & 36s   \\[-2pt]
  $''$	      &  $''$ & $''$       &    $''$    &     $''$    & Ks & 36s   \\[-2pt]
  $''$	      &  $''$ & $''$       & 18:11:22.0 & $-$31:41:49 & J  & 36s   \\[-2pt]
  $''$	      &  $''$ & $''$       &    $''$    &     $''$    & H  & 36s   \\[-2pt]
  $''$	      &  $''$ & $''$       &    $''$    &     $''$    & Ks & 36s   \\[-2pt]
  $''$	      &  $''$ & $''$       & 18:11:05.0 & $-$31:41:49 & J  & 36s   \\[-2pt]
  $''$	      &  $''$ & $''$       &    $''$    &     $''$    & H  & 36s   \\[-2pt]
  $''$	      &  $''$ & $''$       &    $''$    &     $''$    & Ks & 36s   \\[-2pt]
11-12 Jun 2001 & bulge & SOFI-SMALL & 18:11:05.0 & $-$31:45:49 & J  & 1180s \\[-2pt]
  $''$	      &  $''$ & $''$       &    $''$    &     $''$    & H  & 1530s \\[-2pt]
  $''$	      &  $''$ & $''$       &    $''$    &     $''$    & Ks & 2880s \\[-2pt]
 9-11 Jun 2001 & $''$ & $''$       & 18:11:13.5 & $-$31:45:49 & J  & 1200s \\[-2pt]
  $''$	      &  $''$ & $''$       &    $''$    &     $''$    & H  & 1020s \\[-2pt]
  $''$	      &  $''$ & $''$       &    $''$    &     $''$    & Ks & 1920s \\[-2pt]
 9-11 Jun 2001 & $''$ & $''$       & 18:11:13.5 & $-$31:43:49 & J  & 1200s \\[-2pt]
  $''$	      &  $''$ & $''$       &    $''$    &     $''$    & H  & 1020s \\[-2pt]
  $''$	      &  $''$ & $''$       &    $''$    &     $''$    & Ks & 1920s \\[-2pt]
 9-11 Jun 2001 & $''$ & $''$       & 18:11:05.0 & $-$31:43:49 & J  & 1200s \\[-2pt]
  $''$	      &  $''$ & $''$       &    $''$    &     $''$    & H  & 1020s \\[-2pt]
  $''$	      &  $''$ & $''$       &    $''$    &     $''$    & Ks & 1920s \\[-2pt]
 11 Jun 2001  & disk  & SOFI-LARGE & 19:07:32.0 & $-$05:19:57 & J  & 600s  \\[-2pt]
  $''$	      &  $''$ & $''$       &    $''$    &     $''$    & H  & 510s  \\[-2pt]
  $''$	      &  $''$ & $''$       &    $''$    &     $''$    & Ks & 960s  \\[-2pt]
15 Apr 1999   & bulge &  WFI       & 18:10:17.0 & $-$31:45:16 & V  &  20s  \\[-2pt]
  $''$	      &  $''$ &  WFI       &    $''$    &     $''$    & V  & 2$\times$300s  \\[-2pt]
  $''$	      &  $''$ &  WFI       &    $''$    &     $''$    & I  &  20s  \\[-2pt]
  $''$	      &  $''$ &  WFI       &    $''$    &     $''$    & I  & 3$\times$300s  \\[-2pt]
\noalign{\smallskip}
\hline
\end{tabular}
\label{logobs}
\end{center}
\end{table*}

Exposures of a disk control field located at ($l,b$)=(30,0)
(RA=19:07:32, DEC=$-$05:19:57) were also obtained for one pointing
of the large field camera in order to estimate the disk
contribution to the bulge CMD (Section~\ref{cmd}).  

Each of the deep exposures, both for the disk and the bulge field, was
obtained with detector setup DIT=6 NDIT=5 (i.e, every frame was the
average of 5 exposures of 6 sec. each), while for the shallow ones we
used DIT=1.2 and NDIT=5\footnote{See the SOFI User Manual for a more
detailed explanation of the instrument setups:\\
\noindent
{\tt http://www.eso.org/lasilla/Telescopes/NEWNTT/sofi}}.  Both of
them were repeated with a random dithering pattern until reaching the
exposure time listed in Table~\ref{logobs}.  Both sky transparency and
seeing were somewhat variable during the run, and for this reason the
total exposure times on different fields were adjusted in order to
compensate for these effects. The typical seeing during the
observations was $0\farcs80\pm0\farcs15$.

Images were pre-reduced using standard IRAF routines. A sky image,
obtained by median combination of the dithered frames of each filter,
was subtracted from each frame. Flat fielding was then performed using
the "SpecialDomeFlat" template which applies the appropriate
illumination corrections, as described in the SOFI User
Manual. Finally, all the dithered frames obtained in a sequence
were averaged in a single image for each filter. In what follows we
will call ``frame'' the combination of each of these sets.

Standard photometry, including PSF modeling, was carried out on each
frame using the DAOPHOT{\small II} photometry package (Stetson
1987). We used all the stars identified in each frame to obtain the
coordinate transformations among the frames.  These transformations
were used to register the frames and obtain a median image.  The
latter, having the highest S/N, was used to create the star list, by
means of a complete run of DAOPHOT{\small II} and ALLSTAR. The final
star list, together with the coordinate transformations, was finally
used as input for ALLFRAME (Stetson 1994), for the simultaneous
PSF-fitting photometry of all the frames of each field.

Only one of the four half--nights assigned to this programme was
photometric. During that night we observed several standards from the
Persson et al. (1998) catalog, three of which just before and after
one of the bulge fields observed with SOFI-SMALL. This field was
calibrated by means of the standard stars, whose frames were
pre-processed in the same way as the science ones. Aperture photometry
within a radius of $5.8$ arcsec was obtained for the standard stars,
and aperture corrections determined using some bulge isolated
and unsaturated stars were applied to the PSF instrumental magnitudes
of the bulge field. The calibration equations obtained from the
standard stars were then applied to the magnitude of the bulge stars,
neglecting the color term due to the very small color range of the
near-IR standard stars.  Having calibrated one bulge field, all the
others were registered to the same photometric system by comparison of
the common stars: a single large field included the whole mosaic of
four small fields, and there was always $\sim 20\%$ overlap among the
four fields of each mosaic.

\begin{figure}
\centerline{\psfig{figure=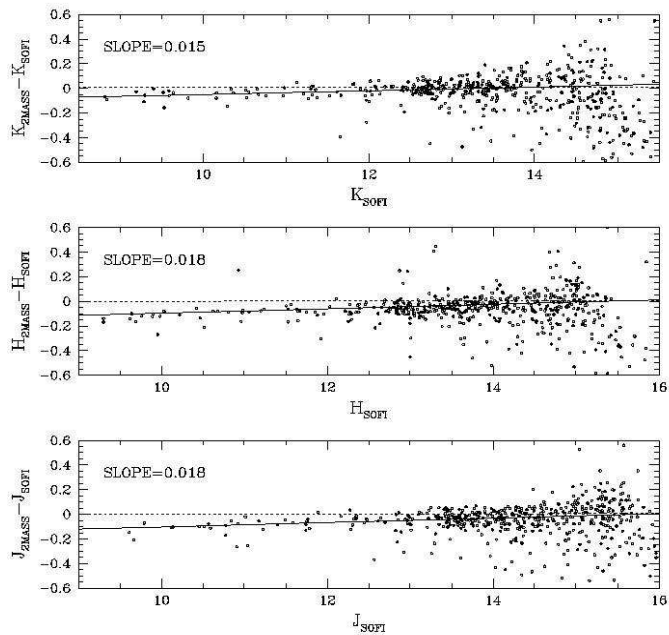,width=9cm}}
\caption{Comparison between the SOFI photometry and that from the 2MASS
survey, for the stars in common. The solid line is the least square fit
to the data, while the dotted one is the relation found by Carpenter
(2001). }
\label{2mass}
\end{figure}

The whole bulge area mapped with SOFI is also included in the 2MASS
near-IR survey (Carpenter 2001), whose second incremental release is
publicly available on the WEB\footnote{ {\tt
http://irsa.ipac.caltech.edu/applications/2MASS}}.  Figure~\ref{2mass}
shows the comparison between the $J,H,K_s$ magnitudes from our
calibrated photometry and the same quantities for the stars in common
with the 2MASS point source catalog.  The latter has a limit magnitude
of $K_s\sim 15$, therefore only the stars measured in one of the four
SOFI-LARGE fields (i.e., in the short exposures) are plotted in this
figure. The solid line represents a least square fit to the data,
while the dotted one is the relation between the 2MASS and the Persson
et al. (1998) photometric system (Carpenter 2001). The latter is based
on 2MASS observations of 82 standard stars from Persson's catalog.

A trend with magnitude is clearly visible in Fig.~\ref{2mass}, with
the difference between the two photometries being zero at the faintest
limit but increasing up to $\sim 0.1$ mag at the brightest end in $J$
and $H$, while remaining quite small ($\lsim 0.05$ mag) in $K_s$.
Such a behavior points to a non-linearity effect in one of the two
detectors. To our knowledge neither the 2MASS nor the SOFI
measurements should be affected by non-linearity; in particular the
SOFI detector has been tested to be linear within $2\%$ up to 10,000
ADU (cf. the SOFI User Manual), and only the three brightest stars in
this plot have counts above this limit.  We are then left with no
explanation for the trend seen in Fig.~\ref{2mass}, but none of our
conclusions relies on an accuracy better than $\sim 0.1$ in the
magnitudes. Moreover, if the error is in our measurements and not in
the 2MASS ones, the trend is identical in the $J$ and $H$ bands,
therefore it would affect the magnitudes but not the $J-H$ color we
used to construct the CMD shown in Section~\ref{sec_cmd}.

The disk control field was not observed during the photometric night,
therefore the only way we had to calibrate its CMD was to rely on
the comparison with 2MASS. However, given that our main
interest was to put the disk stars in the same photometric system
as the bulge ones, we determined the calibration transformations
between our disk instrumental magnitudes and 2MASS, but then also
applied the small differences shown in Fig.~\ref{2mass}, in order
to try and keep consistency between the disk and bulge calibrated
photometry.

\begin{figure}
\centerline{\psfig{figure=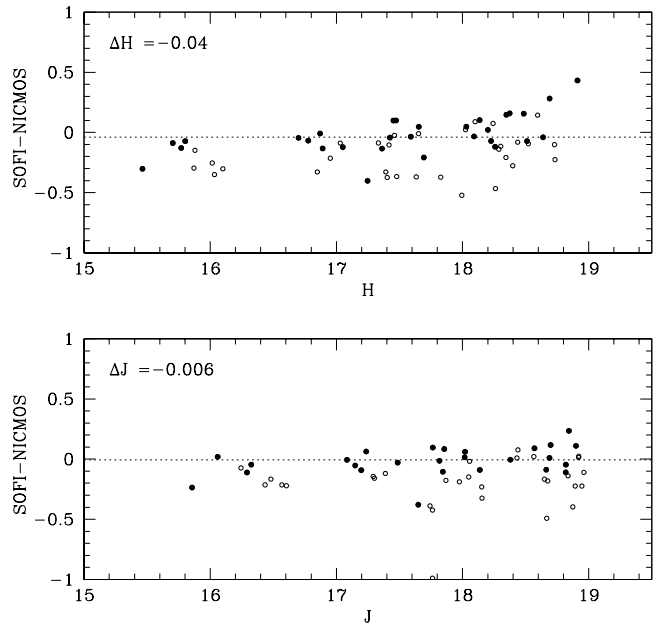,width=9cm}}
\caption{Comparison between the SOFI photometry presented here and the
deep NICMOS photometry from Paper II, for the stars in
common. Open circles refer to stars that may be affected by blending in
SOFI frames.}
\label{nicmos}
\end{figure}

One of the goals of this project is to obtain a complete luminosity
function for the galactic bulge, from the RGB tip down to the faintest
stars ($0.15 M_\odot$) measured with NICMOS. Therefore, it is crucial
to make sure that the NICMOS photometry is consistent with that
obtained here with SOFI, especially given the large uncertainties in
the NICMOS calibration (Stephens et al. 2000).  Figure~\ref{nicmos}
shows the differences in $J$ and $H$ for the stars in common between
the two data sets.  Given the large difference in spatial resolution
between SOFI and NICMOS, this comparison may be affected by
systematics due to the presence of close pairs of stars resolved by
NICMOS but not by SOFI. For this reason, different symbols (filled)
were adopted in Figure~\ref{nicmos} for the stars whose positions
matched within 0.5 SOFI pixels.  These stars are less likely to be
blends, because the presence of a companion {\it resolved by NICMOS}
would likely result in a displacement of the star centroid. The median
differences, in $J$ and $H$, between the common stars, if only the
{\it bona fide} single stars are considered, are $\Delta J=-0.006$ and
$\Delta H=-0.04$ as shown in the figure labels.  If all the common
stars were considered, the median differences would be anyway
negligible for our purposes: $\Delta J= \Delta H=-0.09$.

Completeness estimates were determined via artificial-star
experiments. A total of about 3500 stars were added to the original
SOFI frames, both for the bulge and for the disk fields, with
magnitudes and colors consistent with the RGB+MS instrumental fiducial
lines. In order to avoid to artificially increase the crowding, at the
same time optimizing the CPU time, in each independent experiment the
artificial stars were added along the corners of an hexagonal grid, as
explained in Paper II. As usual, photometry of the artificial frames
was performed following the same procedures as for the original ones.

\begin{figure}
\centerline{\psfig{figure=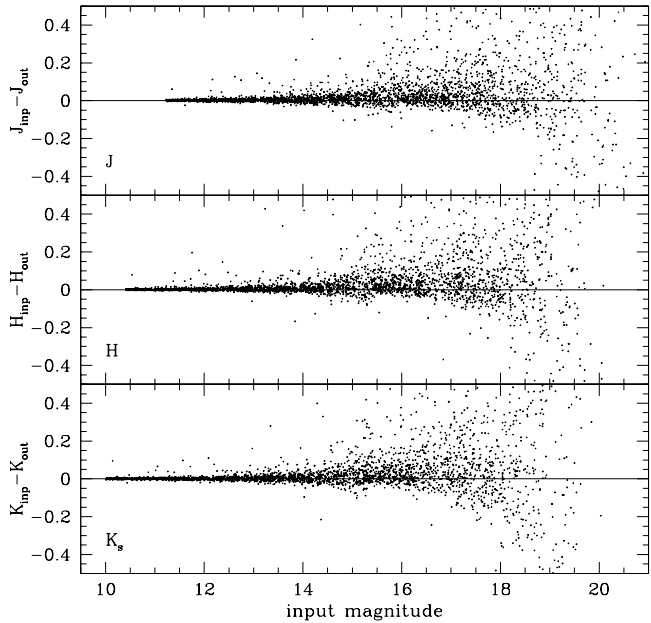,width=9cm}}
\caption{Difference between the input and the output magnitude of
the artificial star experiments on the SOFI frames. The results
obtained for the SOFI-LARGE and SOFI-SMALL frames were matched at
$J=16$ as in the photometry of the real stars.}
\label{crowd}
\end{figure}

Figure~\ref{crowd} shows the results of these experiments, as
the difference between the input and output magnitude of the
artificial stars in each filter.

Note that the distribution is asymmetric about the zero, as occasional
blendings result in brighter output magnitudes compared to the input 
ones. However, the ridge line (i.e., the peak of the distribution)
remains close to zero up to $J\sim18$, and therefore the estimate
of the turnoff magnitude, from the CMD ridge line is not affected by
blending. We expect instead the luminosity function to be affected
(see Section 5.1). The simulations have shown that the SOFI photometry
is more than $50\%$ complete above $J\sim19$ and $H=K_s\sim18$.

\subsection{Optical data} 

A $34\times33$ arcmin bulge field including the whole area mapped with
SOFI was observed with the wide field imager WFI@2.2m telescope, as
part of the EIS PRE-FLAMES programme (Momany et al. 2001). The $V$ and
$I$ raw frames have been retrieved from the public ESO archive, while
the pre-processed images were not released yet.  The frames were taken
under very good seeing conditions, with a PSF FWHM of $0\secondip6$,
as measured on the frames.  Details about the observations are given
in Table~\ref{logobs}. Science images were de-biased and flatfielded
by means of standard IRAF routines, using a set of sky flat-fields
taken the same night.

The photometry for each of the 8 WFI chips was performed separately,
in order to properly model the PSF variation across them.  As usual,
complete photometry and PSF modeling were carried out on each of the
$I$ and $V$ frames, and then coordinate transformations were
calculated among them in order to obtain a median frame to use for a
more complete star finding. The star list obtained in this way was
then used as input for ALLFRAME, which performed simultaneous PSF
fitting photometry on the 3 $V$ and 4 $I$ frames. The whole procedure
was repeated for each of the 8 chips.

Calibration to the Johnson photometric system was performed by means of
a set of Landolt (1992) standard fields observed the same night
through the 8 WFI chips.  The number of standard stars present in
each of these fields has been recently increased by Stetson (2000),
allowing us to measure $\sim 40$ standard stars per chip.
A zero point and a color term were determined separately for each chip.
The color terms were then averaged, and new zero points were
calculated imposing the former to be fixed. Variations $\lsim 0.1$
mag have been found among the photometric zero points of different
chips.

Completeness estimates were determined for the WFI data in the same
way described above for the SOFI ones. WFI data are more than
$50\%$ complete above $I\sim22.5$, and $V\sim23$.

\section{The Color Magnitude Diagram}
\label{sec_cmd}

\subsection{Near-IR CMD} 

\begin{figure}
\centerline{\psfig{figure=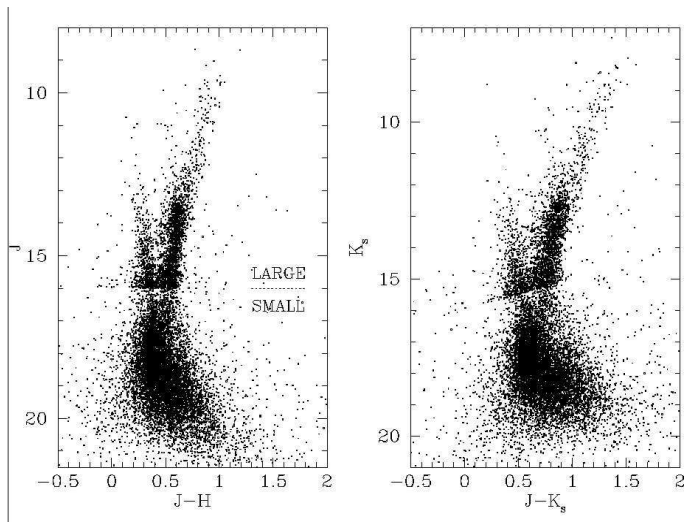,width=9cm,angle=-90}}
\caption{$J,H$ (left) and $J,K_s$ (right) CMD of the bulge stars
imaged with SOFI. In both panels, stars brighter than $J=16$ come from
the $8\farcm3\times8\farcm3$ area mapped with short exposures and the
SOFI-LARGE field, while fainter stars were measured in a $3.9\times3.9$
arcmin area observed with the SOFI-SMALL field.}
\label{cmd}
\end{figure}

Figure~\ref{cmd} shows the near-IR CMD of all the stars measured in
the bulge field.  The jump in the number of stars at $J=16$ is due to
stars brighter than $J=16$ having been sampled from the
$8\farcm3\times8\farcm3$ SOFI-LARGE field, while fainter stars have
been measured with the larger angular resolution and deeper exposures
of the SOFI-SMALL field, which mapped a $3\farcm9\times3\farcm9$ area.

The bulge HB red clump is visible at $J\sim14$ and $J-H=0.6$,
partially merged into the RGB. The large magnitude spread of the HB is
due to a combination of differential reddening, metallicity dispersion
and depth effect. These factors  also cause the HB clump to merge
vertically with the RGB {\it bump}, expected to be located $\sim 0.7$
magnitudes fainter in $J$. The RGB bump (Iben 1968; Rood 1972; Salaris,
Cassisi \& Weiss 2002) is
predicted to be very populated in a high metallicity system as the
bulge, and would have itself a large magnitude spread due to the same
depth and reddening effects mentioned for the HB clump.

The bulge turnoff is clearly visible at $J\sim17.5$ and $J-H\sim0.4$,
while the almost vertical sequence departing from near the bulge
turnoff and extending upwards and bluewards is due to foreground main
sequence stars belonging to the disk, widely dispersed along the line
of sight (Ng \& Bertelli 1996).

\begin{figure*}[t]
\centerline{\psfig{figure=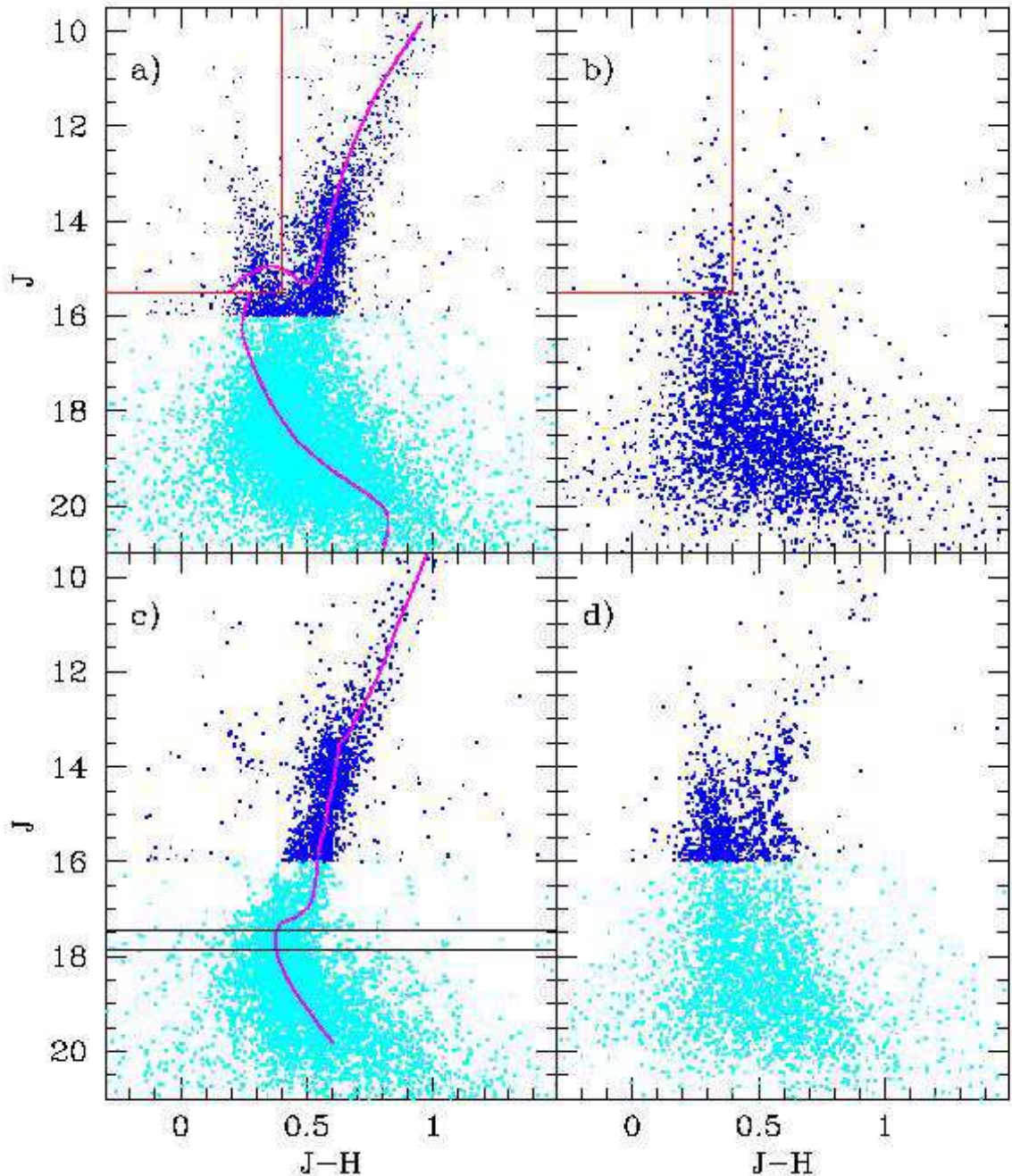,width=17cm}}
\caption{{\it a)} The CMD of the bulge field. The solid line is the 
1~Gyr isochrone for a solar metallicity population. {\it b)} CMD of
the disk control field in the direction $(l,b)=(30,0)$.  The region
inside the box has been used to normalize the number of disk stars
seen through the bulge line of sight.  {\it
c)} CMD of the bulge field as statistically decontaminated from the
disk population.  The horizontal lines locate the main sequence
turnoff at $J=17.65\pm0.2$. The ridge line of the CMD is also 
shown. {\it d)} Stars that were subtracted from the bulge CMD in 
order to obtain the decontaminated CMD.}
\label{decont}
\end{figure*}

Fig.~\ref{decont}a shows the CMD of the bulge field with superimposed
a 1~Gyr isochrone, for a solar metallicity population, adopting a
distance modulus of $(m-M)_0=14.47$ and a reddening of $E(B-V)=0.45$
(see below). Clearly, a population of young stars physically located
inside the bulge, would be definitely bluer than the vertical sequence
of disk stars.  The disk field, $30^\circ$ away from the Galactic
center, was used to statistically decontaminate from the foreground
disk stars the bulge CMD shown in Fig.~\ref{decont}a. Small differences in
the disk stellar population may be expected between the two lines of
sight. Yet, the results have shown that the procedure is indeed quite
effective.  The CMD of the disk control field for an area of 24 square
arcmin is shown in Fig.~\ref{decont}b. The difference in reddening
between the two fields was compensated by shifting the disk CMD shown
in Fig.~\ref{decont}b by 0.17 mag in $J-H$ so as to match the color
location of the blue disk sequences, and in magnitude by the
corresponding $A_J=3.06E(J-H)$ extinction (Cardelli, Clayton \& Mathis
1989).  A region of the CMD free of bulge stars was
then selected in order to normalize the number of disk stars observed
in the disk field to the number of disk stars contaminating the bulge
field.  This region is indicated by the box in the upper left region
of the CMD in Fig~\ref{decont}a.  This normalization is correct only
if no bulge star is present in the box.  

There are 91 stars in the box of the disk control field, which is 4.5
times less than in the corresponding box of the bulge SOFI-LARGE
field.  Therefore, 4.5 stars have been subtracted from the brighter
region ($J<16$) of the bulge CMD for each star seen in the disk CMD.
Since the fainter part of the CMD ($J>16$) was derived from the
SOFI-SMALL field, the 4.5 scaling factor was divided by the ratio of
the SOFI-LARGE to SMALL field area (4.6). Hence 0.98 stars have been
subtracted from the fainter part of the CMD for each star in the disk
control field.

For each disk star in the disk CMD (Fig.~\ref{decont}b) we picked up the
{\it closest} star in the bulge CMD (see below), and subtracted it
according to the normalization factors given above, and to the
slightly different completeness of the two fields. The {\it distance}
on the CMD from a disk star to each bulge star was defined as:
\[
d=\sqrt{[7\times\Delta(J-H)]^2 + \Delta J^2}.
\]
The color difference has been enhanced by a factor of 7 because the
color is much less sensitive than the magnitude to physical
differences (in the distance, reddeninig or mass) between a given disk
star in the control field, and another disk star along the bulge line
of sight. The resulting, clean CMD of the bulge is shown in
Fig.~\ref{decont}c, while the CMD of the stars statistically removed
from the bulge CMD is shown in Fig.~\ref{decont}d. Also shown in this
panel are the fiducial ridge line, helping the eye to identify the
mean branches of the bulge CMD, and the brighter and fainter limits of
our turnoff magnitude estimate ($J=17.65\pm0.2$).

The clumpy appearance of the RGB of Fig.~\ref{decont}d is due to the
poor statistics in this region of the disk CMD since for each disk
star we had to subtract 4.5 times more stars in the bulge field. Hence
for each disk star a small ``cluster'' of close stars was subtracted
from the bulge CMD.  In the attempt to minimize this effect we
actually subtracted every {\it other three} closest stars in the
region $J<16$. We expect a negligible effect on the 0.25 magnitude
bins of the luminosity function discussed in the next Section.

\begin{figure*}[t]
\centerline{\psfig{figure=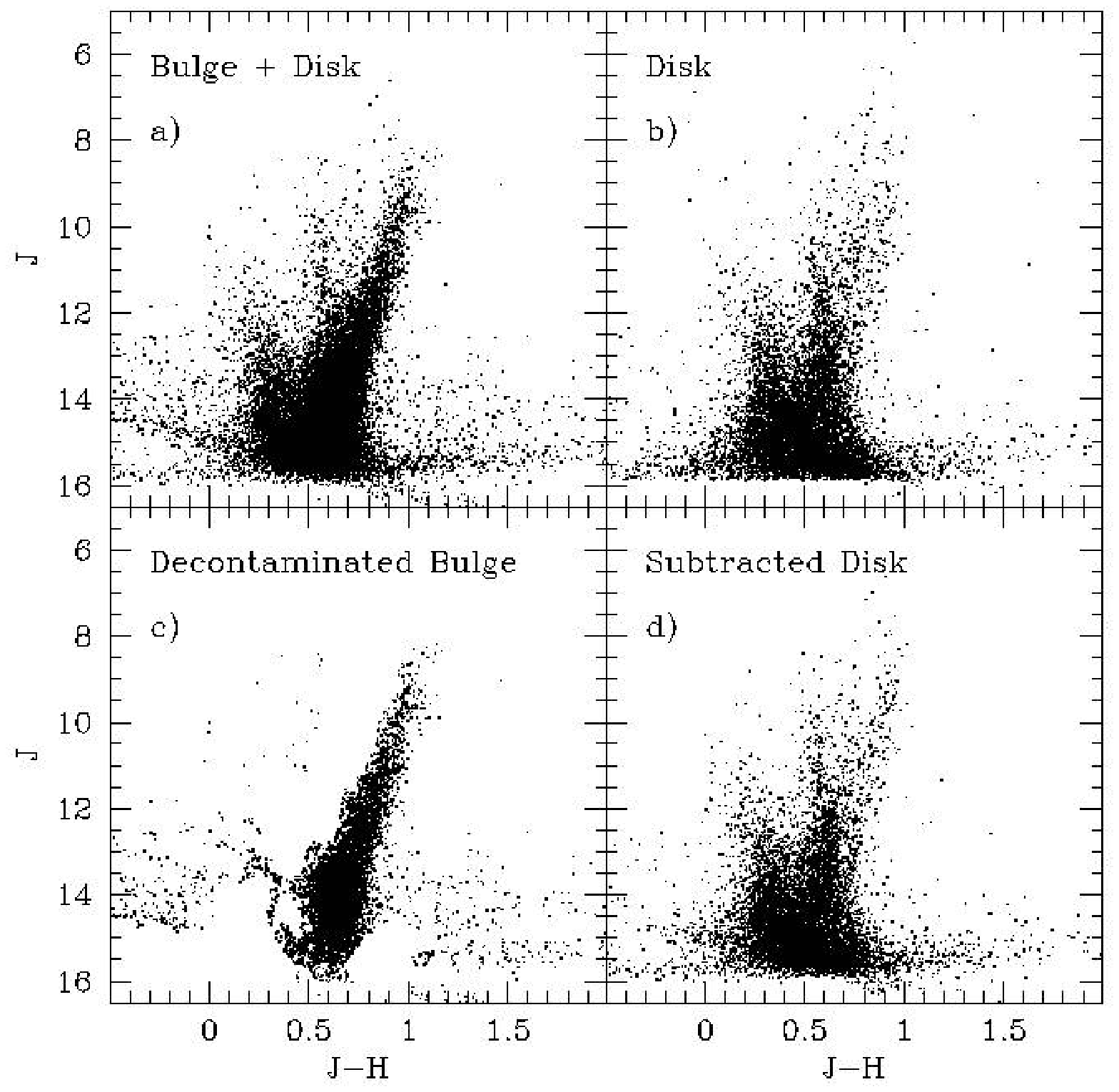,width=16cm}}
\caption{CMDs of the bulge (panel {\it a)} and disk ({\it b)}) fields as
constructed from the 2MASS photometry. {\it c)} The bulge CMD after statistical
decontamination from disk stars; {\it d)} {\it bona fide} disk stars
subtracted from {\it a)} in order to obtain {\it c)}.}
\label{2masscmd}
\end{figure*}

The 2MASS catalog was used to improve the statistics for the bright
part of the CMD and of the luminosity function. Note however, that
the $\sim 4''$ resolution of 2MASS is significantly worse than either
the SOFI or WFI data.  This complicates somewhat the cross
identifications with the SOFI and WFI databases, and crowding effects
are much more severe since two objects closer than $\sim 4''$ are
counted as one in the catalog.

\begin{figure*}[t]
\centerline{\psfig{figure=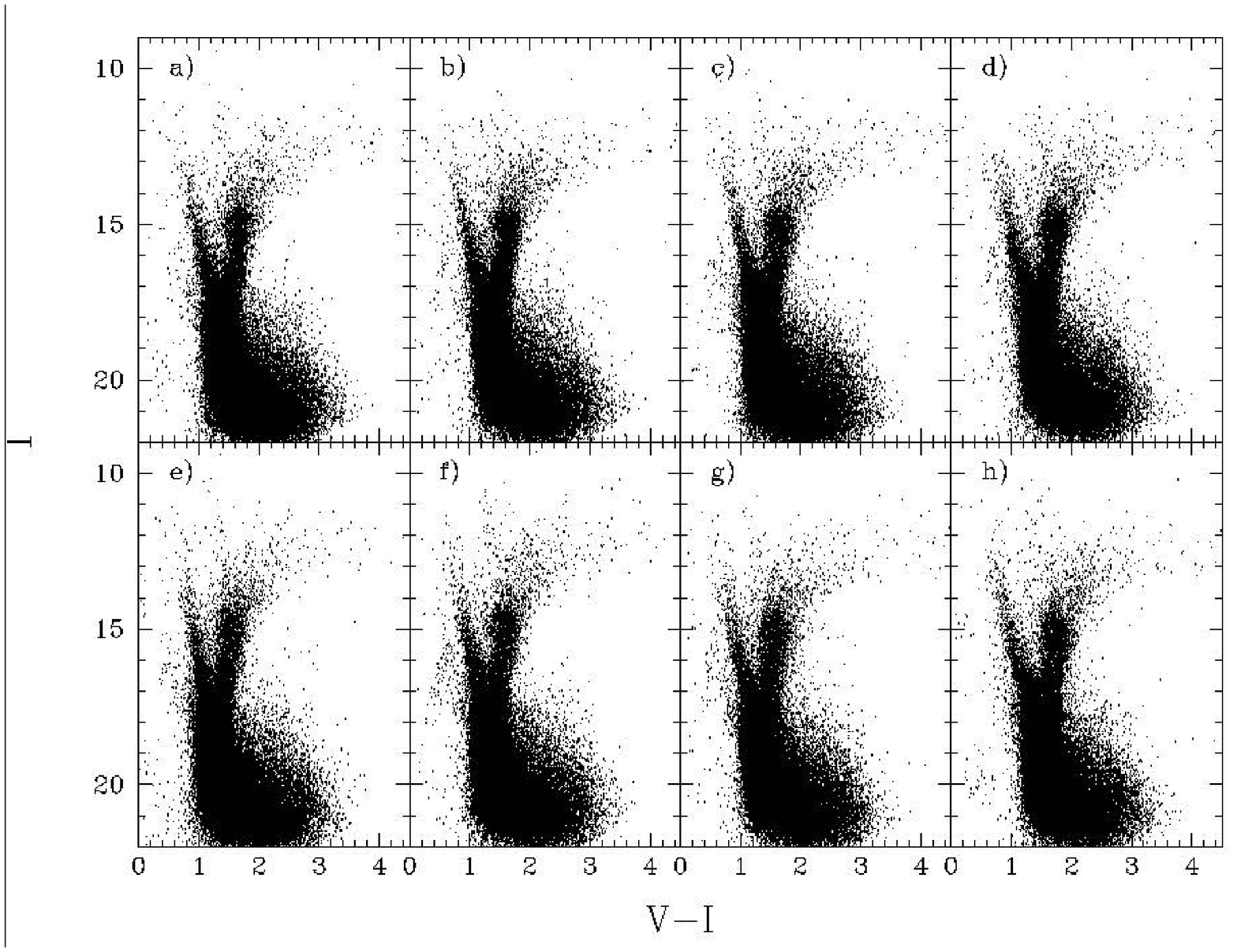,width=16cm,angle=-90}}
\caption{The optical CMD of the bulge from the WFI data. Each panel
corresponds to each of the 8K$\times$8K CCD chips of the WFI camera.}
\label{wfi}
\end{figure*}

From the 2MASS point source catalog we extracted near-IR data for two
large areas, $31\times31$ arcmin each, containing our bulge and disk
field, respectively. A procedure identical to that described above for
the SOFI data has been applied to the bulge CMD of Fig.~\ref{2masscmd}a
in order to statistically decontaminate bulge from disk stars, using
the disk CMD shown in Fig.~\ref{2masscmd}b. Note that since the disk
and bulge field areas are now identical, the scaling factor for the
decontamination (determined within a box similar to that shown in
Fig.~\ref{decont}) is now very close to 1.

The result is again very satisfactory (Fig.~\ref{2masscmd}c,d) for
stars brighter than $J\sim13$.  For fainter stars the decontamination
is less effective because the magnitude and color spread of the disk
and bulge CMD in Fig.~\ref{2masscmd} are quite different, due to the
different drivers of the magnitude limits.  Indeed, due to the low
spatial resolution of the 2MASS data, the photometry is limited by the
crowding in the bulge field, and by the background noise (including
sky) in the disk.  However, in the following only the brightest stars
($J\lsim12$) of the 2MASS decontaminated bulge CMD will be used.

\subsection{The Optical CMD} 

Figure~\ref{wfi} shows the ($I,V-I$) CMD of the 883,417 stars measured
in the $34\times33$ arcmin field of WFI@2.2m. The eight panels in
this figure correspond to the eight CCD chips of the WFI mosaic,
mantaining their relative spatial position. The huge number of points
per panel saturates the plot in the most populated areas, like the
bulge main sequence ($I>18$). The bulge turnoff is located
around $I=18$, while the main sequence of the foreground disk hits the
bulge locus approximately midway between the turnoff and the base of
the RGB.  The HB red clump (merged with the RGB bump) is
visible at $I\sim14.5$ and $V-I\sim1.8$, while the upper RGB (brighter
than the HB) is extremely wide due to the bulge metallicity
dispersion. Note, however, that the narrow sequence departing from the
bulge HB and extending upward, almost vertically is due to the red
clump of the disk stars, dispersed in magnitude as a result of their
large spread in distance (and reddening).  The bulge main sequence
extends almost vertically in this plot, becoming very broad towards
faint magnitudes for the combination of photometric error and plot
saturation, but its increasing skewness towards red colors is in fact
due to the presence of the faint extension of the disk main
sequence.

Although there is evidence for a bulge blue HB population in all the
CMDs (with $V-I\simeq 0.5\div 1$ and $I\simeq 15\div 18$), note that chip\#6
(panel f) contains the bulge globular cluster NGC~6558, already known
to have a blue HB (Rich et al. 1998). The fact that the blue HB seen
in this CMD belongs to the cluster becomes very clear when imposing a
spatial selection around the cluster center. The other cluster
features are not distinguishable from the bulge ones in this plot. A
circular region of 2.4 arcmin radius, centered on NGC~6558, has been
excluded from the following analysis. The radial trend of both cluster
star counts and surface brightness flattens to the background level
well inside this radius, ensuring a negligible cluster contamination
outside 2.4 arcmin.

\begin{figure}
\centerline{\psfig{figure=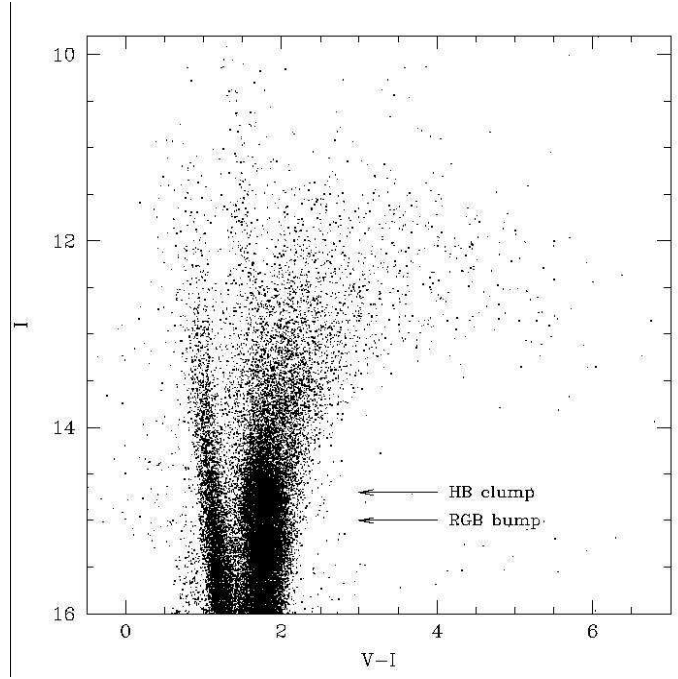,width=9cm}}
\caption{The optical CMD for the bright stars combining the data in the whole
WFI field. The location of the HB clump and the RGB bump is indicated.}
\label{WFIbright}
\end{figure}

For a better visibility of the bright portion of the CMD,
Fig.~\ref{WFIbright} displays only stars brighter than $I=16$, this
time combining data from all the 8 chips of WFI. The disk main
sequence is very prominent on the left side of the diagram, and is
paralleled some $\sim 0.8$ mag redder in $V-I$ by the core helium
burning {\it clump} sequence belonging to the same population.
The bulge RGB and AGB occupy the right side of the diagram, and become very
dispersed in their upper part due to the metallicity dispersion. The
HB clump and the RGB bump are also indicated in the figure.

\section{The Metallicity Distribution of the Bulge}
\label{sec_mdf}

\begin{table}
\begin{center}
\caption[4]{Globular Cluster used as RGB Templates}
\begin{tabular}{cccccc}
\hline
\noalign{\smallskip}
Cluster & [M/H] & $(m-M)_0$ & $E(B-V)$  \\
\noalign{\smallskip}
\hline
\noalign{\smallskip}
NGC 6528  &  --0.10  &  14.37  &  0.62  \\
NGC 6553  &  --0.10  &  13.46  &  0.84  \\
NGC 104   &  --0.57  &  13.32  &  0.05  \\
NGC 6171  &  --0.88  &  13.95  &  0.31  \\
NGC 6121  &  --1.06  &  11.68  &  0.35  \\
NGC 6809  &  --1.59  &  13.82  &  0.10  \\
NGC 7099  &  --1.84  &  14.71  &  0.01  \\
NGC 4590  &  --1.90  &  15.14  &  0.04  \\
NGC 7078  &  --1.92  &  15.15  &  0.09  \\
\noalign{\smallskip}
\hline
\end{tabular}
\label{templates}
\end{center}
\end{table}

\begin{figure}[h]
\centerline{\psfig{figure=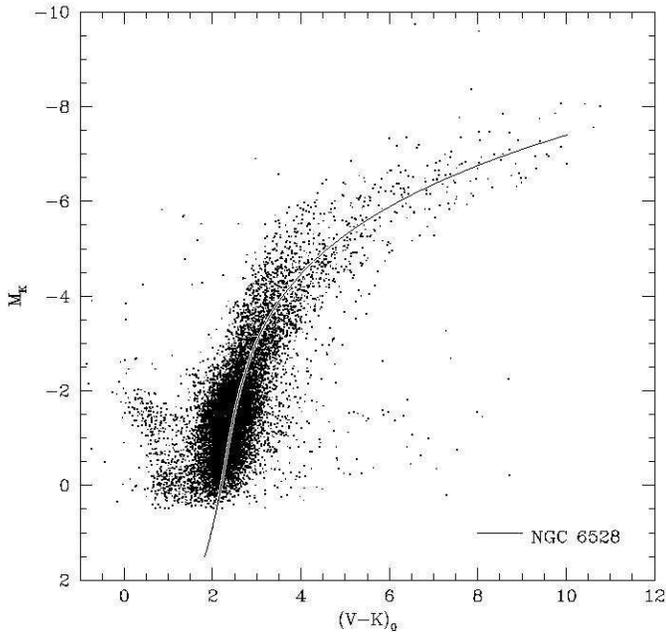,width=9cm}}
\caption{Bulge CMD resulting from the combination of the optical WFI and
near-IR 2MASS data. The line represents the fiducial locus of the
globular cluster NGC~6528. The  distance and reddening of the bulge field
adopted for this comparison are $(m-M)_0=14.47$ and $E(B-V)=0.45$,
respectively, while for NGC~6528 it is assumed $(m-M)_0=14.37$ and
$E(B-V)=0.62$ (Ferraro et al. 2000).  }
\label{n6528}
\end{figure}

From the combination of the optical (WFI) and near-IR (2MASS)
decontaminated data, a ($M_K,V-K$) CMD was obtained including $\sim
22,000$ stars with $M_K<0.5$, and the result is shown in
Fig.~\ref{n6528}. A comparison with the fiducial line of the globular
cluster NGC~6528 from Ferraro et al. (2000) immediately suggests that
the mean metallicity of the bulge is virtually identical to that of
this cluster, with a quite modest dispersion about this mean.  This is
demonstrated by the slope of the giant branch of the cluster and of
that of the bulge field being the same. (Note that the metallicity
affects the {\it slope}, while the reddening causes a solid shift of
the RGB.)  Therefore, the metallicity of NGC~6528 (and that of the
twin cluster NGC~6553) has a special importance in connection with our
attempt at determining the bulge metallicity distribution. In
Fig.~\ref{n6528}, a distance modulus of $(m-M)_0=14.47\pm0.08$ (7.8
kpc) was adopted for the bulge, according to the most recent
determination by McNamara et al. (2000) using RR-Lyrae and $\delta$
Scuti variables from the OGLE survey. The reddening of our bulge field
relative to NGC 6528 can then be estimated from Fig.~\ref{n6528}, and
we obtain $E(B-V)=0.45$, needed to match the bulge and the cluster
loci given the small difference in the two distance moduli.  For
NGC~6528 the corresponding quantities are given in
Table~\ref{templates}.  The large number of stars in the upper RGB of
Fig.~\ref{n6528}, coupled with the high sensitivity of the $(V-K)$
color to metallicity, and its very small sensitivity to a possible age
dispersion, allows a determination of the bulge MD via the method
described in Saviane et al. (2000).

\subsection{The Method} 

The method is based on the construction of a family of hyperbolas in
the plane ($M_K,V-K$) suitable to represent a grid of upper RGBs from
empirical template globular clusters, with known
metallicities. Each hyperbola is represented by the expression:

\begin{equation}
M_K= a + b\times (V-K) + \frac{c}{(V-K)-d}
\end{equation}

\noindent
where the coefficients $a,b,c$ and $d$ are quadratic functions of the
metallicity:

\begin{equation}
a=k_1\mbox{[M/H]}^2 + k_2\mbox{[M/H]} + k_3
\end{equation}
\begin{equation}
b=k_4\mbox{[M/H]}^2 + k_5\mbox{[M/H]} + k_6
\end{equation}
\begin{equation}
c=k_7\mbox{[M/H]}^2 + k_8\mbox{[M/H]} + k_9
\end{equation}
\begin{equation}
d=k_{10}.
\end{equation}

\noindent
Finally, the inversion of equation (1) gives a value of the metallicity
for {\it each point} in the ($M_K,V-K$) plane, hence for each bulge
star.

\begin{figure}
\centerline{\psfig{figure=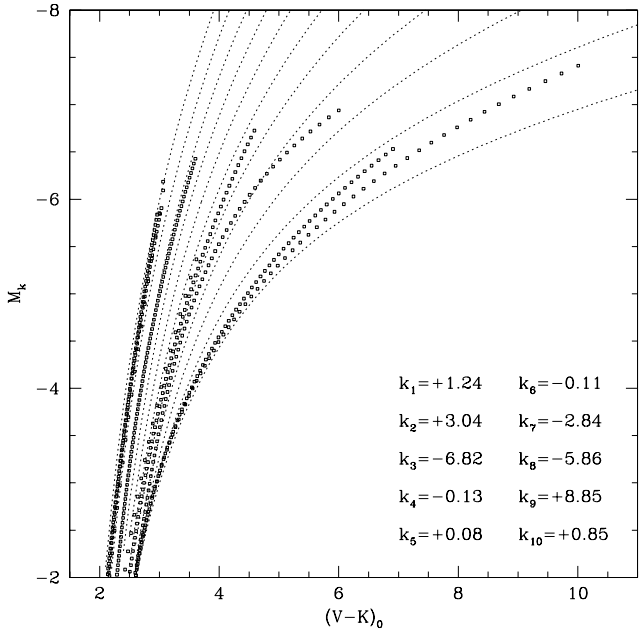,width=9cm}}
\caption{The $V,K$ RGBs of the template GCs published
by Ferraro et al. (2000) (open squares), together with the analytical
RGBs (dotted lines). The values of [M/H] for the analytical RGBs go
from solar to [M/H]=$-2.0$ in steps of 0.2.}
\label{rgbs}
\end{figure}

Although in principle spectroscopic determinations may be more
accurate, the photometric approach has the advantage of relying on
high S/N quantities (i.e., the magnitudes of the brightest stars) and
on large statistics. However, it is entirely differential, i.e., it
depends on the accuracy of the metallicity assigned to the templates,
which ultimately relies on spectroscopic determinations. The most
complete and homogeneous grid of near-IR GC RGB templates presently
available in the literature (Ferraro et al.\ 2000; see
Table~\ref{templates}) includes 10 low reddening, nearby GCs with
metallicity [M/H] between $-1.92$ and $-0.1$. One cluster (NGC~6637)
from the Ferraro et al. sample has been excluded from our
calculations, for reasons explained below.

With the exception of the two most metal rich clusters (NGC 6528 and
NGC 6553), for the GCs in Table~\ref{templates} we adopted the iron
abundance [Fe/H] from Harris (1996), and corrected it to [M/H]
adopting an $\alpha$-element enhancement [$\alpha$/Fe]=0.3 for
clusters with [Fe/H]$<-1.0$, and [$\alpha$/Fe]=0.20 for more
metal-rich clusters (Carney 1996; Salaris \& Cassisi 1996). As
emphasized earlier, the case of NGC 6528 and NGC 6553 is particularly
important. A few stars in each of the two clusters have been observed
at intermediate and high spectral resolution, but different groups
have obtained quite discrepant results. For NGC 6528, Carretta et
al. (2001) and Coelho et al.  (2001) report respectively [Fe/H]=+0.07
and $-0.5$ (the latter value coming from low-resolution spectra). For
[M/H] the same authors derive +0.17 and $-0.25$, respectively. For NGC
6553 Barbuy et al. (1999) give [Fe/H]=$-0.55$ and [M/H]=$-0.08$, while
Cohen et al.  (1999) report [Fe/H]=$-0.16$, and Origlia, Rich \&
Castro (2002) give [Fe/H]=$-0.3$, with [$\alpha$/Fe]=+0.3. These
discrepancies are uncomfortably large, and may hopefully disappear as
soon as more high-resolution data are gathered at 8-10m class
telescopes. For both clusters we finally adopt [M/H]=$-0.1$ (the value
reported in Table~2) with the uncertainty of the resulting MD being
dominated by the uncertainty of the metallicity of these two clusters.

Figure~\ref{rgbs} shows the resulting grid of RGB loci: open squares
represent the fiducial points extracted from the empirical templates,
while the RGBs from the analytical interpolation are shown as dotted
lines. The coefficients of the Equations 4--7 are listed inside the
figure.  The analytical RGBs are shown for metallicities ranging from
[M/H]=0 to [M/H]=$-2.0$, in steps of 0.2. All the empirical templates,
with the exception of NGC~104 (third cluster from the right) follow
very well the RGB shape trend of the analytical solution. We could
find no obvious explanation for the apparent discrepancy of NGC~104:
its CMD is quite populated in this region, and, being one of the best
studied clusters, the adopted parameters (metallicity, reddening and
distance) are rather robust. The near-IR observations of this cluster
were done with the ESO/MPG 2.2m telescope mounting IRAC-2, an early IR
array detector (Ferraro et al. 2000); perhaps the use of a more modern
instrument may clarify the issue.

As evident from Figure~\ref{rgbs}, the sensitivity to metallicity is
a strong function of the position in the CMD, hence of metallicity itself.
The leverage increases with both luminosity and color, being favored
at high metallicities by the increasing TiO blanketing in the $V$ band
causing redder and redder $V-K$ colors.

\begin{figure}
\centerline{\psfig{figure=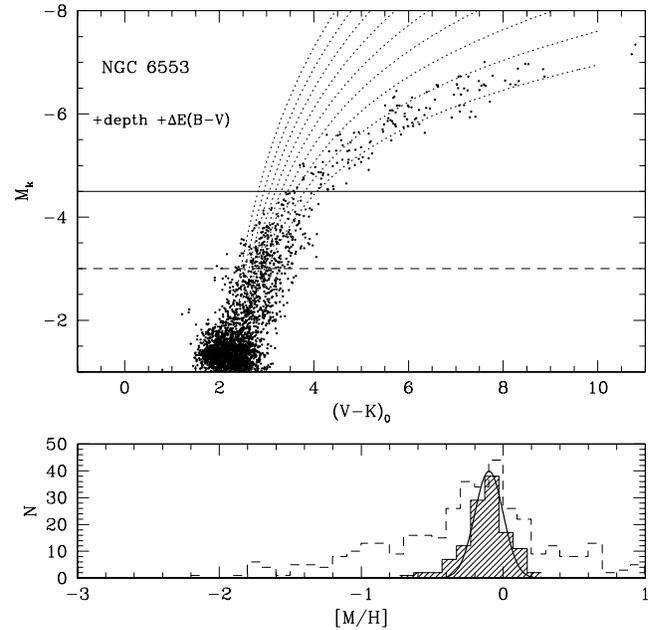,width=9cm}}
\caption{Upper panel: the CMD of the globular cluster NGC~6553 with 
overplotted the analytical RGB templates. The dispersion in distance
typical of the bulge field at $b=-6^\circ$ and a reddening dispersion
of $\Delta E(B-V)=0.3$ have been added to the cluster CMD in order to
simulate the observational biases affecting the derived bulge
MD. Lower panel: MD obtained for NGC~6553.  Shaded histogram is the MD
obtained if only the stars brighter than $M_K=-4.5$ are
considered. This MD is well represented by a Gaussian centered in
[M/H]=$-0.1$ and $\sigma=0.1$.}
\label{n6553}
\end{figure}

In order to investigate the effect of distance and reddening
dispersion on the MD derived using the adopted method we simulated
such effects on the ($M_K,V-K$) CMD of the bulge cluster NGC~6553
(Guarnieri et al.\ 1998). To each star of NGC~6553 we associated a
distance modulus randomly extracted according to a bulge space density
distribution of the form $\rho={\rm Const}\times r^{-3.7}$ (Sellwood
\& Sanders 1988; Terndrup 1988), where $r$ is the spatial distance
from the Galactic center. The line of sight integration at the field
position ($(l,b)=(0,-6)$) gives a 1-$\sigma$ dispersion of 0.13 mag in
the distance modulus, the distribution being close to Gaussian in the
core, but with somewhat broader wings. A top hat reddening
distribution of $\pm 0.15$ was also adopted. The latter is believed to
be an upper limit to the observed bulge reddening variation across our
field, but has been chosen as a conservative assumption, in order to
have an upper limit on the possible biases in the derived MD. Random
extractions of both distance modulus and reddening were repeated 6
times for each star in the original CMD by Guarnieri et al. (1998) in
order to obtain a more populated CMD. Figure~\ref{n6553} shows the
resulting simulated CMD including  distance and reddening dispersion
effects (upper
panel).  The lower panel shows the MD obtained in the same way as that
to be applied to the bulge, and reveals two interesting effects. First,
if all the stars brighter than $M_K=-3$ are considered, the resulting
MD (dashed histogram) is extremely broad with a long metal-poor
tail. This is entirely due to the fact mentioned above: the
separation between the RGBs corresponding to different metallicities
becomes very small towards faint magnitudes, much smaller than the
magnitude/color dispersion due to distance and  reddening
dispersions.  However, if the analysis is restricted to stars brighter
than $M_K=-4.5$ then the derived MD shows a fairly sharp peak at
[M/H]=$-0.1$ (i.e., centered on the adopted metallicity for
this cluster), well fitted by a Gaussian distribution with $\sigma=0.1$
(FWHM=0.24).  This exercise demonstrates that, as long as the analysis
is confined to the stars brighter than $M_K=-4.5$: {\it i}) one should
not expect systematic biases; {\it ii}) the spread introduced by the
distance and reddening dispersions is of the order of $\sim 0.1$ dex
in [M/H]; and {\it iii}) contamination by AGB stars (which would
introduce a bias artificially skewing the distribution towards the
metal-poor side) is negligible at these magnitudes due to the
exclusion of the more populated (and bluer) AGB ``clump'', located at
$M_K\sim-3$. We therefore restrict to $M_K<-4.5$ the derivation of the
bulge MD.

\subsection{The Resulting Metallicity Distribution}

\begin{figure}
\centerline{\psfig{figure=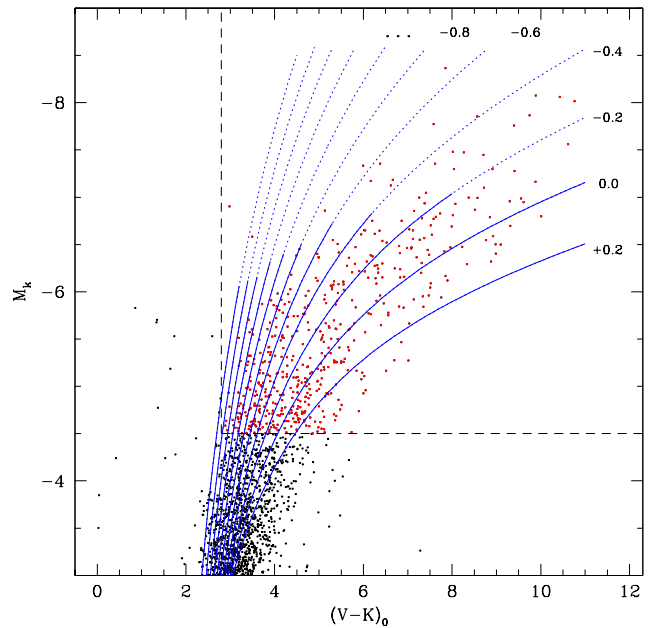,width=9cm}}
\caption{The bulge $V,K$ CMD compared with the analytical RGB templates.
The adopted bulge distance modulus is $(m-M)_0=14.47$ while the mean
reddening is $E(B-V)=0.45$. Only the stars inside the dashed box where
used for the MD determination. }
\label{MDF_mk}
\end{figure}

Figure~\ref{MDF_mk} shows the analytical RGBs overplotted to the bulge
upper RGB. The dotted part of each RGB is the extrapolation above the
theoretical location of the RGB tip (c.f., Section~7), hence only AGB
stars are expected in this region. The
metallicities of the analytical RGBs range from [M/H]=+0.2 to $-1.8$,
in steps of 0.2. Note that the most metal-rich templates have
metallicity of [M/H]=$-0.1$, therefore the shape of the analytical
RGBs for higher metallicity is the result of a small extrapolation.  Stars
bluer than $V-K=2.8$ (vertical dashed line) were not considered, in
order to exclude a few bluer objects that are most likely a disk
residue due to imperfect statistical decontamination. Stars fainter
than $M_K=-4.5$ (horizontal dashed line) were excluded as well, for
the reasons explained above.  After these cuts, a sample of 503 stars
has been used to derive the bulge MD. It is already clear from this
figure that most of the bulge stars are located between the [M/H]$\simeq
0.1$ and [M/H]=$-0.6$ templates, with a peak at [M/H]=$-0.1$ and very few
stars more metal-poor than $\sim -1$. Formally, super-solar
metallicity stars ([M/H]$>0$) represent 29$\%$ of the sample. This
should be regarded as an upper limit, due to the {\it diffusion} to
higher [M/H] values caused by the dispersion in distance, reddening,
etc. (see Fig.~\ref{n6553}).

\begin{figure}
\centerline{\psfig{figure=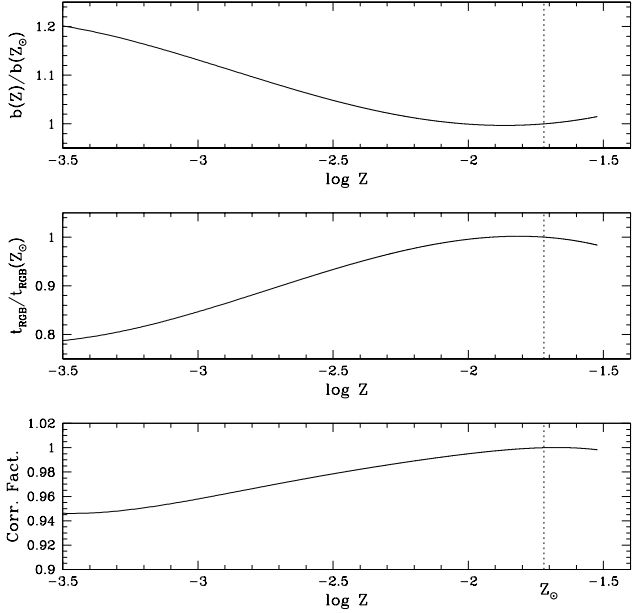,width=9cm}}
\caption{Upper panel: the ratio of the evolutionary flux at any $Z$
to the same quantity at solar $Z$. Middle panel: the time spent on the
RGB from $M_K=-4.5$ to the tip, with respect to the value at
$Z_\odot$. Lower panel: the product of the two factors above, i.e.,
the correction factor that one should apply to the MD found from
the CMD, in order to take into account these effects.}
\label{btnew}
\end{figure}

Before proceeding to derive the bulge MD a few biases must be
quantitatively evaluated and compensated for: 1) the rate at which
stars leave the main sequence (called {\it evolutionary flux} in
Renzini \& Buzzoni 1986) has a slight metallicity dependence, 2) the
rate of evolution along the RGB scales as $Z^{-0.04}$ (Rood 1972), and
3) by adopting the $M_K=-4.5$ cut one samples (bolometrically)
deeper on the RGB the higher the metallicities.

\begin{figure}
\centerline{\psfig{figure=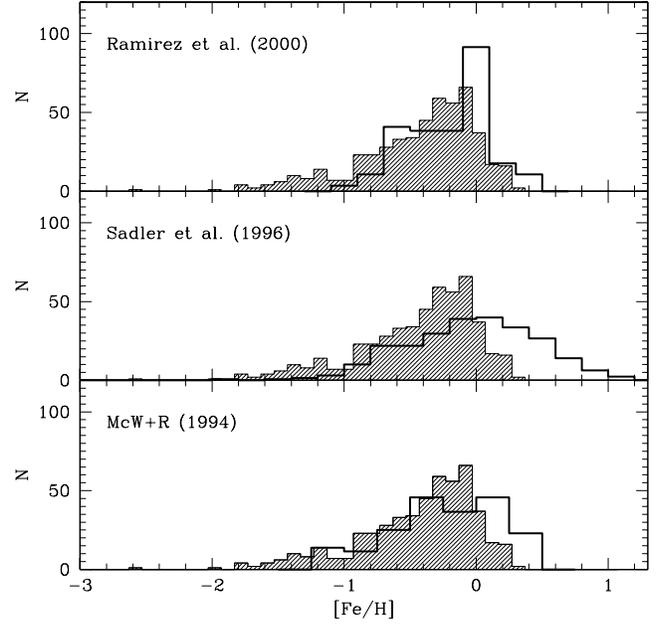,width=9cm}}
\caption{Comparison between the bulge MD derived from Fig.~\ref{MDF_mk} (shaded
histogram) and those derived with spectroscopic surveys.}
\label{MDF}
\end{figure}

The amplitude of these factors is shown in Fig.~\ref{btnew}. The upper
panel shows how the evolutionary flux changes as a function on
metallicity, relative to its value at $Z_\odot$. The
middle panel shows the combination of effect 2) and 3) mentioned
above: the time spent on the RGB, from $M_K=-4.5$ to the tip,
normalized at solar value, slowly increases as a function of metallicity.
Finally the lower panel shows the combination of the two factors
above, which is the correction to apply to each metallicity
bin of the MD to take into account the evolutionary effects.
Since the evolutionary flux and the RGB lifetime have opposite
behaviour with metallicity, the final correction factor is never
larger than few percent, well within the intrinsic precision of
our measurement.

The resulting bulge MD for the 503 stars in Fig.~\ref{MDF_mk} is
finally shown in the three panels of Fig.~\ref{MDF} as a shaded
histogram.  Also shown in Fig.~\ref{MDF}, as thick ``open''
histograms, are the bulge MDs determined from various spectroscopic
studies, normalized to the total number of stars in our sample.  Note
that all these spectroscopic determinations of the bulge MD refer to
[Fe/H] abundance. Since the template RGBs are on the [M/H] scale, we
subtracted from the bulge [M/H] distribution the same $\alpha$-element
enhancement (i.e., 0.2 for [Fe/H]$>-1$ and 0.3 for [Fe/H]$<-1$) that
had been applied to the template GCs, in order to obtain a [Fe/H]
distribution.  Although the bulge high resolution studies do not
permit to draw strong conclusions on the $\alpha$-element enhancement,
the similarity of the bulge CMD with that of bulge GCs would favor a
similar chemical enrichment history.  Certainly the new generation
multi-object spectrographs will allow us to improve our knowledge in
this field.

The comparisons shown in Fig.~\ref{MDF} shows that the present,
``photometric'' MD is broadly consistent with the spectroscopic ones
by McWilliam \& Rich (1994) and Ramirez et al. (2000), with just a
somewhat less pronounced supersolar [Fe/H] tail in the photometric
MD. (This characteristic is exacerbated in the comparison with the MD
of Sadler et al. (1996), based on low resolution data.)  We caution,
however, that the position of the high [Fe/H] cutoff is entirely
dependent on the metallicity assigned to the two template clusters
NGC~6528 and NGC~6553. For instance, the apparent discrepancy with,
e.g., McWilliam \& Rich (1994) at high metallicity would disappear if
we would have assigned to NGC~6528 and NGC~6553 a metallicity $\sim
0.2$ dex higher. Indeed this would bring to coincidence with M\&R the
high metallicity cutoff, while stretching the distribution and
reducing the excess at [Fe/H]$\sim -0.2$.  We believe that the
homogeneous high-resolution studies of cluster {\it and} bulge field
stars that will soon become available, e.g., with the forthcoming
FLAMES multifibre spectrograph on VLT (Pasquini et al. 2000), will
clarify this problem.

On the other hand, we tend to consider as quite reliable the sharpness
of the cutoff at high [M/H].  Due to the TiO blanketing in the $V$
band, the $V-K$ color of RGB stars has a strong, rapidly increasing
sensitivity to [M/H], and any high metallicity tail would have
resulted in a dramatic broadening of the upper RGB, which instead is
clearly absent in the CMD of the bulge shown in
Fig.~\ref{n6528}. Moreover, the high [Fe/H] cutoff would be even
sharper, had we underestimated the effect of the distance and
reddening dispersions.

\subsection{Inferences for the Chemical Evolution of the Bulge}

\begin{figure}
\centerline{\psfig{figure=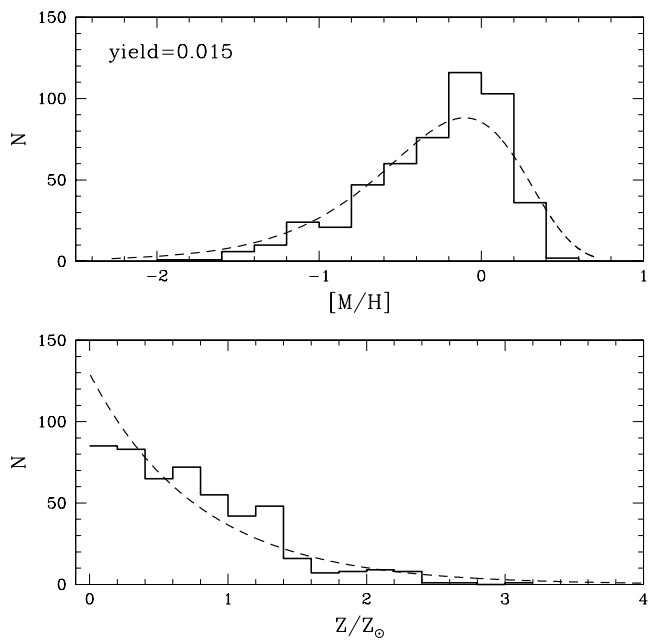,width=9cm}}
\caption{Comparison between the metallicity distribution of bulge
giants from the present study (histogram) and the prediction of the
Simple Model ({\it closed-box} model) with the indicated value of the
yield ($y\equiv \langle Z \rangle$). The model distribution has been
scaled by the total area under the data histogram.  There is a clear
shortage of stars near zero metallicity, and an excess of stars near
solar abundance, but the one-zone model appears to fit the abundance
distribution reasonably well.}
\label{closed}
\end{figure}

The most straightforward approach is now to compare this empirical MD
with the Simple (or one-zone, closed-box) Model of chemical evolution
(Searle \& Sargent 1972).  Rich (1990) first made this comparison for
the bulge and found the Simple Model to be a good fit to the abundance
distribution from Rich (1988).  The Simple Model assumes that no
infall or outflow of metals has occurred during the star-forming phase
of the system.  The distribution follows the relation $N(Z)=y^{-1}{\rm
exp}^{-Z/y}$, where the single parameter $y$, the yield, is given by
$y=\langle Z\rangle$.  Hence, the yield is the average metallicity of
the stars in a one-zone system following the exhaustion of the gas
(Hartwick 1976; Mould 1984).  If metals are lost from the system by
outflow, the functional form of the distribution is unchanged, but the
mean abundance $\langle Z\rangle$ is the yield (as would be calculated
from supernova models) reduced by the appropriate outflow parameter.

Fig.~\ref{closed} shows a plot of the Simple Model distribution (for
$y=\langle Z\rangle = 0.015$) overlaying the data.  Note that both
panels show the same model and the same data, once using [M/H] and
once using $Z/Z_\odot$ as abscissa.  As the Simple Model satisfies the
requirements of a probability distribution, the function is scaled by
the area under the data histogram, so that the area covered by data
and fit are identical.

From the plots, the general shape of the abundance distribution is in
fairly good agreement with the Simple Model. However, there are
noticeable deviations: in the subsolar regime ($Z< 0.3\, Z_\odot$)
less stars are observed than predicted, while with respect to the
model there is an excess of near-solar metallicity objects.This excess
can be traced back to the fairly sharp clustering of the bulge stars
around the RGB template at such metallicity (see again Fig. 7 and
10). This feature is most likely real, as many factors conjure to
broaden the derived distribution (dispersion along the line of sight,
differential reddening, incomplete disk decontamination, etc.). For
the same reasons, most likely real is the sharp cutoff at high
metallicity, that may even be sharper given the mentioned effects that
tend to smooth out any sharp feature in the MD.  The moderate shortage
of metal-poor stars compared to the Simple Model may flag the presence
of a {\it G-dwarf} problem, a point on which we return in Section 8.

The sharp high-metallicity cutoff of the MD suggests that star
formation did not proceed to complete gas consumption.  If the bulge
formed in a rapid and intense starburst, one can imagine that much of
the metals would have been produced {\it in situ} in core-collapse
supernovae, eventually driving strong, metal-rich galactic
winds. There are at least two independent arguments favoring this
scenario, one direct and one indirect: a) high redshift galaxies
thought to be proto-bulges are observed to have strong metal enriched
winds (Steidel et al. 1996), and b) in clusters of galaxies at least
as much metal mass is out of galaxies in the intracluster medium, as
there is locked into stars inside galaxies, which is taken as evidence
for early metal-enriched galactic winds (Renzini 1997, 2002).
Therefore, a relatively sharp cutoff at the metal rich end, such as we
observe, would result if the star formation was sufficiently violent
as to eventually evacuate the gas from the proto-bulge before it was
exhausted by star formation.

\section{The Luminosity Function of the Bulge}
\label{sec_lf}

\begin{figure}
\centerline{\psfig{figure=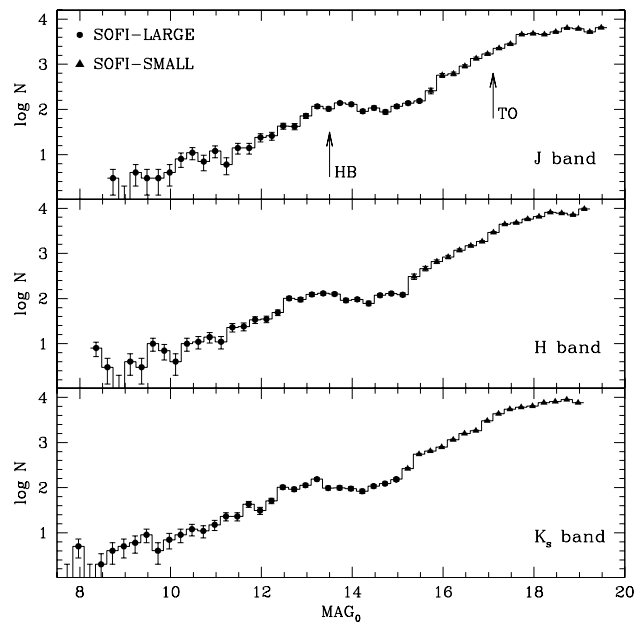,width=9cm}}
\caption{The bulge near-IR LF as obtained from the SOFI data.}
\label{lf}
\end{figure}

The disk-decontaminated CMD shown in Fig.~\ref{decont}c has been used
to construct the bulge luminosity function (LF) from the RGB tip down
to $\sim 1$ mag below the main sequence turnoff, in each of the three
bands: $J,H$ and $K_s$, with the results being shown in Fig.~\ref{lf}.
The number counts in the SOFI-SMALL field ($J>16$, $J_0>15.61$) have been
normalized to those in the SOFI-LARGE field, according to the ratio of
their areas.

For an easier comparison with the LFs of other objects, or different
bulge regions, the LFs presented here are always shown as a function
of dereddened magnitude. The adopted average E$(B-V)$ is 0.45, as
derived from the comparison with the fiducial loci of the CMD of
NGC~6528 (Fig.~\ref{n6528}). The relations between the
absorptions in different bands have been assumed to be $A_V=3.1\times E(B-V)$,
$A_I=0.479\times A_V$, $A_J=0.282\times A_V$,
$A_H=0.190\times A_V$, and $A_K=0.114\times A_V$ (Cardelli
et al. 1989).

The broad peak at $J_0\sim13.2$ in Fig.~\ref{lf} is the HB clump,
slightly ``bimodal'' because merged with the RGB bump, as already
discussed in Section~\ref{cmd}. The main sequence turnoff is located
at $J_0=17.05\pm0.20$, as determined from Fig.~\ref{decont}. The steep
decrease in the number of stars that could be expected just above the
turnoff, is in fact smeared in this LF due to depth effect,
differential reddening, metallicity dispersion, photometric errors
and blending effects (see below).

\begin{figure}
\centerline{\psfig{figure=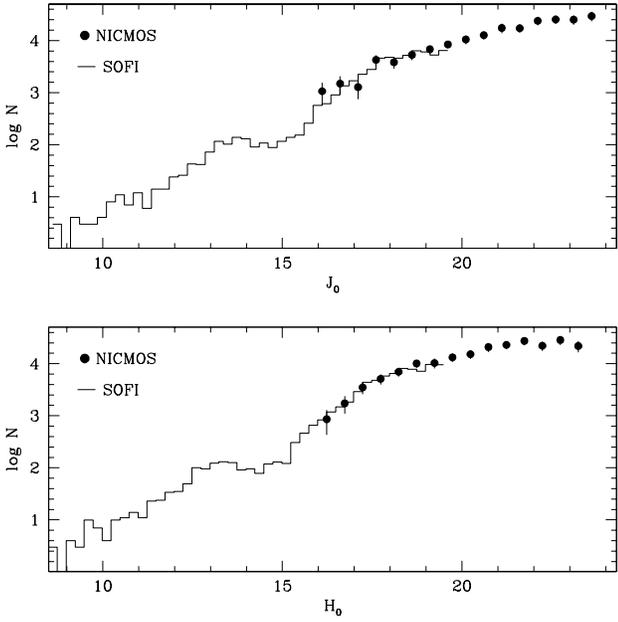,width=9cm}}
\caption{The bulge LF from this work complemented with the one
obtained in Paper II from NICMOS data.}
\label{lfnic}
\end{figure}

Figure~\ref{lfnic} shows the comparison, in $J$ and $H$, between the
SOFI LF and the very deep NICMOS LF from Paper~II.  A simple
normalization by the ratio of the field areas brings the NICMOS LF in
perfect agreement with the SOFI one, based on a field more than 100
times larger. Note that the disk contribution to the NICMOS LF was
subtracted using Kent (1992) model LF of disk and bulge (cf.,
Paper~II for details).

\begin{figure}
\centerline{\psfig{figure=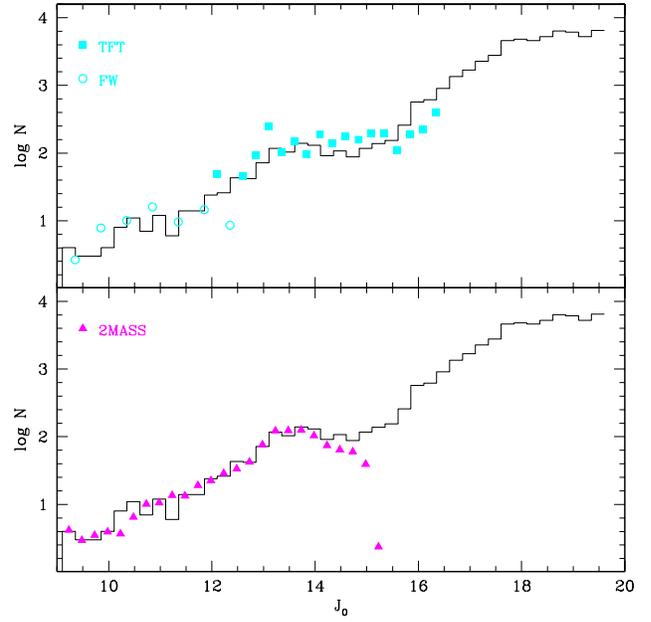,width=9cm}}
\caption{Comparison with previously published bulge LFs. Counts from
different sources have been scaled according to the ratios of the
field area and to the different surface brightness, if referring to
another bulge region. Histogram: this work; filled squares: data from
Tiede, Frogel \& Terndrup (1995); open circles, data from Frogel \&
Whitford (1987); filled triangles: data from the 2MASS sky survey.}
\label{lfother}
\end{figure}

The upper panel of Fig.~\ref{lfother} shows the comparison between the
SOFI LF and the LF by Tiede, Frogel \& Terndrup (1995; TFT) in the
same bulge region, but on a smaller area (4056 arcsec$^2$). The disk
contribution has been subtracted from the TFT star counts using the
ratio between bulge and disk stars computed from Fig.~\ref{decont}a,b. 
The upper panel of Fig.~\ref{lfother} also shows the LF obtained by
Frogel \& Whitford (1987; FW in the figure label) for the M giants in
Baade's Window. The latter has been normalized both for the different
area and the surface brightness difference between the two fields.

The lower panel of Fig.~\ref{lfother} shows the comparison with the
counts from the 2MASS survey, also normalized only for the ratio of
the field areas. The 2MASS counts plotted here were extracted from a
region of 927 arcmin$^2$, centered on the SOFI field.  In the range
where they are complete, i.e., for $J_0<15$ the 2MASS counts agree
perfectly well with the SOFI counts, and are consistent with TFT
and FW counts.

\begin{figure}
\centerline{\psfig{figure=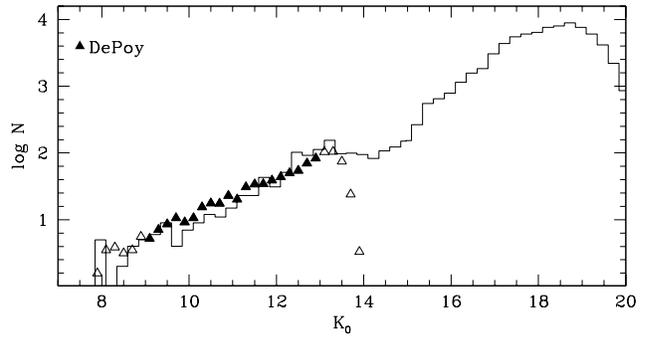,width=9cm}}
\caption{Comparison with the $K$-band LF by DePoy et al. (1993).
Open symbols refer to counts affected by incompleteness at the faint
end and by saturation at the bright end.}
\label{depoy}
\end{figure}

Finally, Fig.~\ref{depoy} shows the comparison with the $K$-band
bulge LF from DePoy et al. (1993). The latter was obtained from 
the photometry of a 604 square arcmin fields towards Baade's 
Window, and included disk stars. The two LF where normalized
according to the different field area and surface brightness, 
and the foreground disk contamination, estimated from the
present data, has been subtracted from the Depoy LF. 

\begin{table}
\begin{center}
\caption[2]{Complete near-IR bulge luminosity function$^{\mathrm{a}}$ }
\begin{tabular}{rrrrrr}
\hline
\noalign{\smallskip}
  $J_0$  & log N$_J$ & $H_0$   & log N$_H$ & $K_0$ & log N$_K$  \\
\noalign{\smallskip}
\hline
\noalign{\smallskip}
  7.98 & -0.242 &  7.61 &  0.117 &  7.27 & -0.008 \\[-2pt]
  8.23 & -0.184 &  7.86 & -0.008 &  7.52 &  0.168 \\[-2pt]
  8.48 & -0.485 &  8.11 &  0.144 &  7.77 &  0.256 \\[-2pt]
  8.73 & -0.133 &  8.36 &  0.293 &  8.02 &  0.418 \\[-2pt]
  8.98 &  0.027 &  8.61 &  0.504 &  8.27 &  0.469 \\[-2pt]
  9.23 &  0.621 &  8.86 &  0.526 &  8.52 &  0.418 \\[-2pt]
  9.48 &  0.469 &  9.11 &  0.536 &  8.77 &  0.481 \\[-2pt]
  9.73 &  0.547 &  9.36 &  0.594 &  9.02 &  0.557 \\[-2pt]
  9.98 &  0.594 &  9.61 &  0.805 &  9.27 &  0.776 \\[-2pt]
 10.23 &  0.566 &  9.86 &  0.794 &  9.52 &  0.800 \\[-2pt]
 10.48 &  0.811 & 10.11 &  0.962 &  9.77 &  0.904 \\[-2pt]
 10.73 &  1.007 & 10.36 &  1.062 & 10.02 &  0.926 \\[-2pt]
 10.98 &  1.027 & 10.61 &  1.138 & 10.27 &  1.144 \\[-2pt]
 11.23 &  1.136 & 10.86 &  1.149 & 10.52 &  1.095 \\[-2pt]
 11.48 &  1.128 & 11.11 &  1.210 & 10.77 &  1.260 \\[-2pt]
 11.73 &  1.279 & 11.36 &  1.352 & 11.02 &  1.250 \\[-2pt]
\und{11.98} & \und{1.352} & 11.61 &  1.383 & 11.27 &  1.428 \\[-2pt]
 12.23 &  1.415 & 11.86 &  1.447 & 11.52 &  1.416 \\[-2pt]
 12.48 &  1.633 & 12.11 &  1.553 & 11.77 &  1.500 \\[-2pt]
 12.73 &  1.623 & 12.36 &  1.733 & 12.02 &  1.568 \\[-2pt]
 12.98 &  1.857 & 12.61 &  2.003 & 12.27 &  1.784 \\[-2pt]
 13.23 &  2.068 & \und{12.86} & \und{2.075} & 12.52 &  2.006 \\[-2pt]
 13.48 &  2.013 & 13.11 &  2.090 & \und{12.77} &  \und{2.085} \\[-2pt]
 13.73 &  2.143 & 13.36 &  2.114 & 13.02 &  2.121 \\[-2pt]
 13.98 &  2.114 & 13.61 &  2.100 & 13.27 &  2.188 \\[-2pt]
 14.23 &  1.959 & 13.86 &  1.959 & 13.52 &  1.991 \\[-2pt]
 14.48 &  2.033 & 14.11 &  1.982 & 13.77 &  1.996 \\[-2pt]
 14.73 &  1.944 & 14.36 &  1.892 & 14.02 &  1.978 \\[-2pt]
 14.98 &  2.068 & 14.61 &  2.072 & 14.27 &  1.919 \\[-2pt]
 15.23 &  2.140 & 14.86 &  2.111 & 14.52 &  2.033 \\[-2pt]
 15.48 &  2.188 & 15.11 &  2.083 & 14.77 &  2.090 \\[-2pt]
 15.73 &  2.412 & 15.36 &  2.487 & 15.02 &  2.179 \\[-2pt]
 15.98 &  2.756 & 15.61 &  2.665 & 15.02 &  2.187 \\[-2pt]
 16.23 &  2.790 & 15.86 &  2.818 & 15.27 &  2.422 \\[-2pt]
 16.48 &  2.957 & 16.11 &  2.918 & 15.52 &  2.740 \\[-2pt]
 16.73 &  3.130 & 16.36 &  3.071 & 15.77 &  2.813 \\[-2pt]
 16.98 &  3.228 & \und{16.61} & \und{3.191} & 16.02 &  2.916 \\[-2pt]
\und{17.23} &\und{3.376} & 16.86 & 3.320 & 16.27  & 3.080 \\[-2pt]
 17.48 &  3.518 & 17.11 &  3.478 & 16.52 &  3.218 \\[-2pt]
 17.73 &  3.666 & 17.36 &  3.598 & 16.77 &  3.285 \\[-2pt]
 17.98 &  3.612 & 17.61 &  3.676 & 17.02 &  3.504 \\[-2pt]
 18.23 &  3.591 & 17.86 &  3.737 & 17.27 &  3.661 \\[-2pt]
 18.48 &  3.676 & 18.11 &  3.803 & 17.52 &  3.760 \\[-2pt]
 18.73 &  3.764 & 18.36 &  3.890 & 17.77 &  3.802 \\[-2pt]
 18.98 &  3.814 & 18.61 &  3.977 & 18.02 &  3.826 \\[-2pt]
 19.23 &  3.854 & 18.86 &  4.013 & 18.27 &  3.904 \\[-2pt]
 19.48 &  3.901 & 19.11 &  4.008 & 18.52 &  3.926 \\[-2pt]
 19.73 &  3.952 & 19.36 &  4.033 & 18.77 &  3.971 \\[-2pt]
 19.98 &  4.001 & 19.61 &  4.094 & \und{19.02} & \und{3.903} \\
 20.23 &  4.039 & 19.86 &  4.138 & & \\[-2pt] 
 20.48 &  4.075 & 20.11 &  4.162 & & \\[-2pt] 
 20.73 &  4.141 & 20.36 &  4.213 & & \\[-2pt] 
 20.98 &  4.219 & 20.61 &  4.290 & & \\[-2pt] 
 21.23 &  4.246 & 20.86 &  4.335 & & \\[-2pt] 
 21.48 &  4.232 & 21.11 &  4.348 & & \\[-2pt] 
 21.73 &  4.262 & 21.36 &  4.385 & & \\[-2pt] 
 21.98 &  4.343 & 21.61 &  4.433 & & \\[-2pt] 
 22.23 &  4.400 & 21.86 &  4.417 & & \\[-2pt] 
 22.48 &  4.410 & 22.11 &  4.356 & & \\[-2pt] 
 22.73 &  4.401 & 22.36 &  4.361 & & \\[-2pt] 
 22.98 &  4.397 & 22.61 &  4.432 & & \\[-2pt] 
 23.23 &  4.413 & 22.86 &  4.452 & & \\[-2pt] 
 23.48 &  4.449 & 23.11 &  4.386 & & \\[-2pt] 
 23.73 &  4.492 & 23.36 &  4.290 & & \\[-2pt] 
 23.98 &  4.535 &       &        & & \\[-2pt] 
\noalign{\smallskip}
\hline
\end{tabular}
\label{lfjhktab}
\begin{list}{}{}
\item[$^{\mathrm{a}}$] Horizontal lines mark the boundaries between 2MASS,
SOFI and NICMOS based data, from top to bottom.
\end{list}
\end{center}
\end{table}

\begin{table}
\begin{center}
\caption[3]{Optical bulge luminosity function.}
\begin{tabular}{rrrr}
\hline
\noalign{\smallskip}
  $V_0$  & log N$_V$ & $I_0$   & log N$_I$ \\
\noalign{\smallskip}
\hline
\noalign{\smallskip}
 11.73 & -1.110 & 10.46 & -0.381 \\
 11.98 & -1.106 & 10.71 & -0.038 \\
 12.23 & -0.406 & 10.96 &  0.501 \\
 12.48 &  0.010 & 11.21 &  0.646 \\
 12.73 &  0.345 & 11.46 &  0.773 \\
 12.98 &  0.663 & 11.71 &  0.918 \\
 13.23 &  0.910 & 11.96 &  1.072 \\
 13.48 &  1.024 & 12.21 &  1.153 \\
 13.73 &  1.204 & 12.46 &  1.269 \\
 13.98 &  1.370 & 12.71 &  1.252 \\
 14.23 &  1.552 & 12.96 &  1.402 \\
 14.48 &  1.707 & 13.21 &  1.498 \\
 14.73 &  1.855 & 13.46 &  1.559 \\
 14.98 &  2.079 & 13.71 &  1.685 \\
 15.23 &  2.138 & 13.96 &  1.950 \\
 15.48 &  2.089 & 14.21 &  2.143 \\
 15.73 &  2.058 & 14.46 &  2.087 \\
 15.98 &  2.133 & 14.71 &  2.090 \\
 16.23 &  2.130 & 14.96 &  1.953 \\
 16.48 &  2.119 & 15.21 &  1.864 \\
 16.73 &  2.240 & 15.46 &  2.001 \\
 16.98 &  2.346 & 15.71 &  2.036 \\
 17.23 &  2.467 & 15.96 &  2.119 \\
 17.48 &  2.718 & 16.21 &  2.213 \\
 17.73 &  2.906 & 16.46 &  2.368 \\
 17.98 &  3.126 & 16.71 &  2.480 \\
 18.23 &  3.232 & 16.96 &  2.685 \\
 18.48 &  3.334 & 17.21 &  2.937 \\
 18.73 &  3.398 & 17.46 &  3.122 \\
 18.98 &  3.408 & 17.71 &  3.226 \\
 19.23 &  3.457 & 17.96 &  3.371 \\
 19.48 &  3.453 & 18.21 &  3.414 \\
 19.73 &  3.454 & 18.46 &  3.500 \\
 19.98 &  3.475 & 18.71 &  3.529 \\
 20.23 &  3.477 & 18.96 &  3.542 \\
 20.48 &  3.466 & 19.21 &  3.583 \\
 20.73 &  3.462 & 19.46 &  3.606 \\
 20.98 &  3.436 & 19.71 &  3.601 \\
 21.23 &  3.457 & 19.96 &  3.611 \\
 21.48 &  3.422 & 20.21 &  3.654 \\
 21.73 &  3.422 & 20.46 &  3.609 \\
 21.98 &  3.456 & 20.71 &  3.614 \\
 22.23 &  3.400 & 20.96 &  3.417 \\
 22.48 &  3.359 & 21.21 &  3.287 \\
\noalign{\smallskip}
\noalign{\smallskip}
\hline
\end{tabular}
\label{lfvitab}
\end{center}
\end{table}

From the combination of the data from 2MASS, SOFI, and NICMOS, a
composite LF for the Galactic bulge can be constructed, using the best
data for each luminosity range. The result is reported in
Table~\ref{lfjhktab}, that lists the star counts for $J$, $H$, and
$K_s$ bands, while Table~\ref{lfjvitab} lists the optical $V$ and $I$
LF.  Note that NICMOS data are available only in the $J$ and $H$
bands.  All the counts have been normalized to the area of
$8\farcm3\times8\farcm3$ mapped by the SOFI-LARGE field: i.e., the
2MASS counts have been divided by 13.43, the SOFI-SMALL counts have
been multiplied by 4.6, and the NICMOS counts have been multiplied by
609. These scaling factors can be used to calculate the Poissonian
errors associated with the counts in each bin. The numbers in
Table~\ref{lfjhktab} have been corrected for both disk contamination
and incompleteness, although the latter is only significant for the
few faintest bins of the NICMOS LF, and it is always $\lsim 50\%$.

\subsection{Simulated CMD and Theoretical LF} 

In order to compare our observations to corresponding theoretical
predictions we have developped a simulator which generates the CMD of
a stellar population with a single age and a wide metallicity
spread. In this way, we neglect the presence of an age spread, which
is justified since the location of the RGBs of (relatively) old
stellar populations is much more sentitive to metallicity than to
age. The code results from a development of the CMD simulator used in
Greggio et al. (1998), adapted to describe single age Stellar
Populations with a wide metallicity spread. A thorough description of
it can be found in Rejkuba (2002, PhD thesis). We give here a short
report, and specify the ingredients used in our simulations.

\begin{figure}
\centerline{\psfig{figure=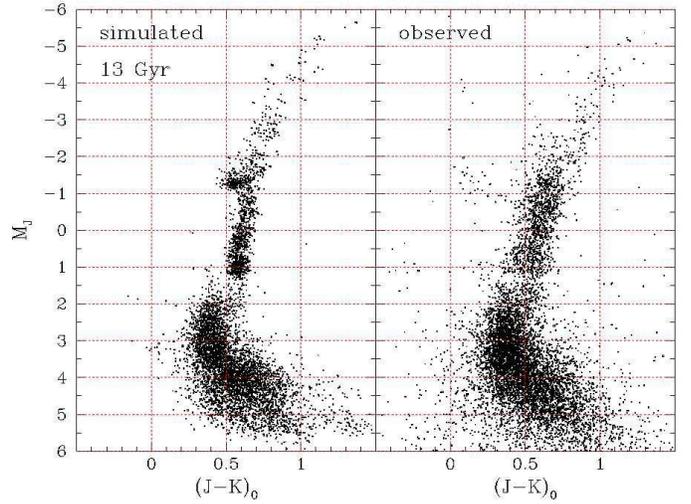,width=9cm,angle=-90}}
\caption{Left panel: simulated CMD for a 13~Gyr old population
with the MD determined in Section~4. Right panel: the bulge CMD
for the SOFI-LARGE field.}
\label{cmdsim}
\end{figure}

A Monte Carlo procedure is used to extract mass and metallicity of a
simulated star, which gets then positioned on the H-R diagram via
interpolation among isochrones. The determined bolometric magnitude
and effective temperature are transformed into monochromatic magnitudes
by interpolating within bolometric correction tables.  Incompleteness
and photometric errors, as measured on the real frames are applied. In
particular this includes the brightening effect due to blending that
was discussed in Section 2.1. The procedure is iterated until the
number of objects observed in some CMD region is reached.

The isochrones data base consists of Cassisi and Salaris (1997)
models, implemented with Bono et al. (1997) , plus some additional
models explicitly computed for this application with the code
described in Cassisi and Salaris (1997) and Cassisi, Salaris and Bono
(2002).  The metallicity range goes from Z=0.0001 to Z=0.04, and the
helium abundance varies in locksteps with Z following $Y\simeq
0.23+2.5Z$. 

Theoretical bolometric corrections (BCs) have been obtained by
convolving the model atmospheres computed by Castelli et al. (1997)
with the Landolt $V$ and $I$ and the SOFI $J,H,K_s$ filter passbands
and fixing the zero point such that $BC_{V}=-0.07$ for the model of
the Sun, and all the colors are zero for the model of Vega. However, the
brightest stars in our sample are cooler than $T\sim4000$K, limit
below which current model atmospheres are known to fail to reproduce
the observed spectra, due to the inappropriate treatment of
molecules. Therefore, the empirical BCs by Montegriffo et al. (1998)
where used instead of the theoretical ones for temperatures
$T\lsim4000$ K.  The dependence of the BCs on the metallicity
parameter is taken into account in the simulations.

The results of the artificial stars experiments described in Section~2
have been used to assign the detection probability and the photometric
error (i.e. the difference between input and output magnitude) to the
simulated stars. The synthetic CMD has thus the same observational
biases as that obtained from the measured frames.

A number of simulations have been computed for ages ranging from
8 to 15 Gyr, and by adopting:
\\
{\it i)} an IMF slope of $-1.33$, as found in Paper II \\
{\it ii)} the metallicity distribution derived in Section~4 \\
{\it iii)} an average distance modulus of $(m-M)_0=14.47$ and
reddening of $E(B-V)=0.42$\footnote{This value is $0.03$ magnitudes
smaller than the one adopted in Section~4, which was derived from the
comparison of the mean locus of the $(M_V,V-K)$ bulge CMD with that of
NGC~6528. The lower value of $E(B-V)=0.42$ is instead required for a
best fit with the adopted isochrones, and therefore for a consistent
comparison with the synthetic CMD.}.
Each obtained synthetic CMD has been further dispersed for depth
(($\rho(r)\propto r^{-3.7}$) and variable reddening
($\Delta E(B-V)=\pm 0.15$) effects.

\begin{figure}
\centerline{\psfig{figure=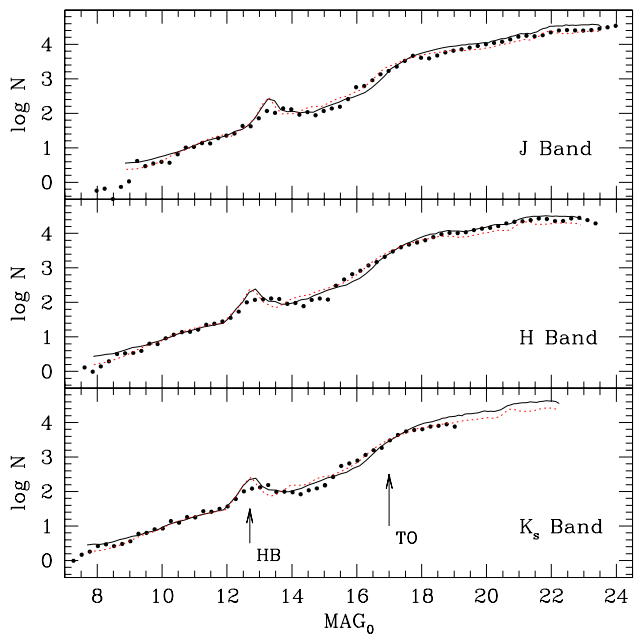,width=9cm}}
\caption{The complete bulge LF in the three near-IR bands (dots),
as resulting from the combination of the 2MASS, SOFI and NICMOS data
(from Table~\ref{lfjhktab}).  A theoretical LF for an age of 13~Gyr
(solid line) and 8~Gyr (dotted line) is shown as a solid line for
comparison.}
\label{lftot}
\end{figure}

Figure~\ref{cmdsim} (left panel) shows the synthetic CMD for a 13~Gyr
old population with the characteristics described above, and a
comparison with the bulge near-IR CMD observed with SOFI is shown in the
right panel.

The simulation code was run until the number of stars with $M_J<1.2$
reached the number of stars sampled by SOFI-LARGE in this CMD
region. Then it was run again until extracting as many stars with
$M_J>1.2$ as in the SOFI-SMALL field.

The main purpose of the comparison in Fig.~\ref{cmdsim} is to show how
much the simulation code is able to reproduce the observed CMD,
including the observational biases introduced by the dispersions in
distance, reddening and metallicity, as well as blending.

Figure~\ref{cmdsim} shows that indeed the main features of the
observed CMD are well reproduced by the simulator. Noticeable
exceptions are the morphology of the HB clump, and of the lower
RGB. In fact, the observed HB is significantly less defined than the
simulated one, and also the color width of the RGB is underestimated
by about a factor of two in $J-K$.  In principle, this mismatch can be
caused by several effect, although none of them, alone, seems
convincing to us. Larger spread in the observed CMD can be due to an
underestimate of one of the following effects:\\

{\it i)} Distance spread: the bulge density law could be {\it flatter}
than $\rho\propto r^{-3.7}$, and/or the bulge being a bar may increase
the distance dispersion along the line of sight. Note that adopting a
larger distance dispersion would smear the HB clump, making is more
similar to the observed one, but would not have any appreciable effect
on the spread in color of the RGB. Hence, a larger distance spread
alone would not be sufficient to make the simulated CMD identical to
the observed one.

{\it ii)} Differential reddening. Larger values of this parameter
would broaden the RGB of the simulated CMD, but also would cause the
HB to appear tilted along the reddening line, which is not seen in the
observed CMD of the bulge, while it is prominent in e.g., NGC 6553
(Zoccali et al. 2001a).  Also, the differential reddening needed to
account for the spread in color seen in the near-IR CMD would imply a
too large spread in the optical CMD. An attempt was made to correct
for differential reddening following the method used for NGC 6553
(Zoccali et al. 2001a), but failed due to the small scale of the
reddening variations across the field. Indeed, in the case of NGC
6553, one finds reddening variations of the order of $\delta E(V-I)\sim 0.1$
on scales of only $\sim 20''$.

{\it iii)} Photometric errors. Larger SOFI-LARGE photometric errors
than adopted in the simulation would certainly help smearing the simulated
RGB and HB respectively in color and magnitude, but can hardly solve both
problems at once. To smear enough the simulated HB the photometric error
should be $\sim 0.3-0.4$ mag, which would produce a too broad RGB compared to
the observed CMD (see  Fig.~\ref{cmdsim}).

{\it iv)} Problems in the theoretical models. The RGB temperatures (hence
colors) are strongly dependent on the mixing-length parameter, which
needs to be empirically calibrated, and there are no perfect
calibrators.  The models used in the present simulations were
calibrated by fitting the RGB of template globular clusters from
Frogel et al. (1983) for [Fe/H]$\lsim -0.7$, and at solar metallicity
by demanding to the solar model to have the solar radius. Ideally, it
would have been preferable to have a homogeneous calibration over the
whole metallicity range, but the solar metallicity clusters would have
introduced in the calibration the uncertainly in their reddening. As a
net result, the adopted mixing length is some $10\%$ larger at [Fe/H]=0
than at [Fe/H]$\lsim -0.7$, which has the effect of compressing
somewhat the relative spacing of the RGBs as a function of
metallicity. This effect goes indeed in the direction of reducing the
color dispersion in the simulated RGBs. Moreover, the simulation is
also affected by the uncertainty in the bolometric corrections and
color-temperature transformations. All in all, Salaris, Cassisi \&
Weiss (2002) estimate an uncertainty of $\sim 0.10-0.15$ mag in the
range of optical colors spanned by theoretical RGBs, which can also be
taken as indicative of the uncertainty in the range of the near-IR
colors.

While the origin of the dicrepancy of Fig.~\ref{cmdsim} is still
partly unclear, we proceed to the construction of the theoretical LF
from the simulated CMD, keeping in mind that the region around the HB
and lower RGB is presently not well reproduced by our simulations.
The same code with the same inputs was then used to generate a much
larger number of stars compared to the simulation shown in
Fig.~\ref{cmdsim}, in order to construct a smooth LF from the upper
RGB down to the limit of the NICMOS photometry and the result is shown
in Fig.~\ref{lftot}. The simulation includes photometric error and
blending effects, and is meant to match the observed LF after
correction for incompleteness.

Figure~\ref{lftot} finally shows the comparison between the observed
$J,H,K_s$ LFs (dots) and the theoretical LFs generated from the
simulated CMD described above (lines). The observed LFs result from
the combination of all the available data, namely 2MASS + SOFI +
NICMOS all scaled to the SOFI-LARGE area, as in Tables~\ref{lfjhktab}.
The simulated LFs refer to 8 and 13~Gyr old populations.  Observed and
theoretical LFs were normalized to the total number of stars with
MAG$_0<12$ in all bands.  There is overall agreement between the
theoretical and the observed LFs, though the HB clump + RGB bump
appears much sharper in the simulation than in the observed LF, as
expected from Fig.~\ref{cmdsim}.

This comparison also shows that formally the 8~Gyr isochrone gives a slightly
better fit to the data above the turnoff, which is the part of the LF
sensitive to age. However, the same effect causing the smearing of the HB
would have also made shallower the drop off of the luminosity function
just above turnoff, hence making the bulge population to appear younger than
it is when comparing to the simulated LFs. Hence, we believe that before
having identified the origin of the additional RGB and HB dispersion the
comparison of simulated and observed LFs cannot set more stringent
constraints on the age of the bulge stars other than being $\sim 10$ Gyr.

\subsection{The Bulge Spectral Energy Distribution} 

\begin{figure}
\centerline{\psfig{figure=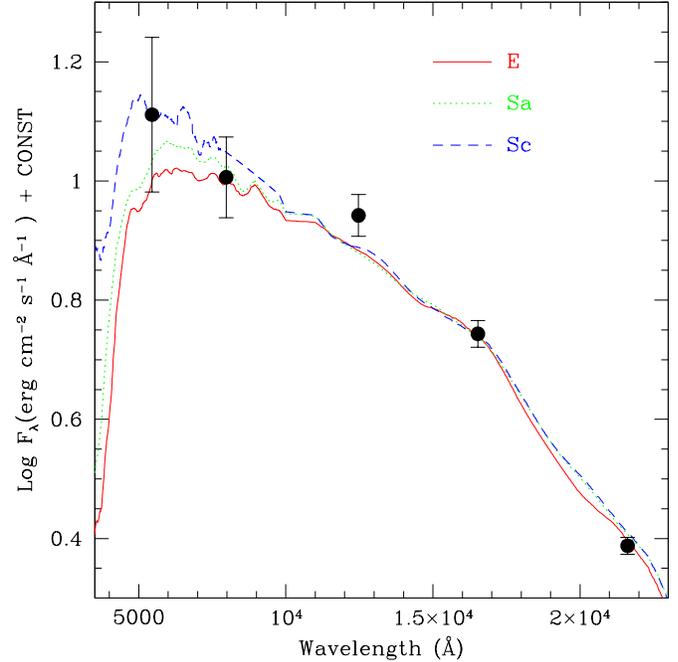,width=10cm}}
\caption{The spectral energy distribution of the bulge summing the
contribution of all individual star (filled circles). The error bars
here show what would be the {\it systematic} displacement of all the
points if an uncertainty of $E(B-V)=\pm 0.1$ is allowed. Superimposed
are the spectral energy distributions of template galaxies from
Mannucci et al. (2001). }
\label{SED}
\end{figure}

Integrating the LFs in Tables~\ref{lfjhktab} and \ref{lfvitab}, one
can determine the total luminosity of the sampled bulge stellar
population in each band. This has been performed using the whole
database (NICMOS, SOFI, 2MASS for the infrared and WFI for the
optical), then normalizing the results to the area of the SOFI-LARGE
field. For the $J$ and $H$ bands the NICMOS data allow to include the
contribution of all stars down to $\sim 0.15 M_\odot$. The $K$-band
contribution of stars fainter than the SOFI limit has been estimated
from the theoretical $M/L$ ratio of lower main sequence stars and
adopting $-1.33$ for the slope of the IMF, and found to be just a fraction
$\sim 10^{-5}$ of the total $K$-band luminosity. This
result is not surprising given the flat bulge IMF.  For the same
reason, we integrated the $V$ and $I$ LFs of Table~\ref{lfvitab}
safely neglecting the contribution of the lower MS stars.  The
resulting sampled luminosities are:
\smallskip\par\noindent
$L_{\rm V}=1.00\times 10^5L_{\rm V,\odot}$\par\noindent
$L_{\rm I}=1.13\times 10^5L_{\rm I,\odot}$\par\noindent
$L_{\rm J}=2.84\times 10^5L_{\rm J,\odot}$\par\noindent
$L_{\rm H}=3.67\times 10^5L_{\rm H,\odot}$\par\noindent
$L_{\rm K}=4.20\times 10^5L_{\rm K,\odot}$,\par\noindent
\smallskip\par\noindent
where the solar magnitudes have been taken from Allen (2000). To
derive monochromatic fluxes, the integrated magnitudes in each band
have been transformed to AB magnitudes, and the monochromatic flux has
been calcualted as log$F_\nu = -0.4\, m_{\rm AB} + 19.44$ (Oke \& Gunn
1983), with $F_\lambda = F_\nu c/\lambda^2$.  Figure~\ref{SED} shows
the resulting spectral energy distribution (SED) of the galactic
bulge, compared to the template spectra of elliptical, Sa and Sc
galaxies from Mannucci et al. (2001). The Galactic bulge follows quite
well the Sc template, which is not surprising given that the Milky Way
is indeed an Sbc galaxy. For the optical part of the spectrum, the
Mannucci et al.  templates rely on those by Kinney et al. (1996), which in
the case of Sc galaxies  come from the
integrated spectra of a region $10''\times 20''$ wide at the center of
the two galaxies NGC~2403 and NGC~598, whose radii are 10.6 and 35.2
arcmin respectively, i.e., the sampled regions are located well inside
the bulge of these galaxies.  Visual inspection of a DSS image of
NGC~2403 suggests that the aperture used to derive the spectrum of
this galaxy almost certainly includes a few site of active star
formation. Their effect is however minimal in the $V$ and $I$ bands,
while dominating the spectrum only in the IUE ultraviolet.

On the other hand, the average value of the Mg$_2$ index of the
ellipticals in the Kinney et al. sample is 0.314, which compares to
Mg$_2$=0.23 (Puzia et al. 2002) for the Galactic
bulge. Hence, our bulge is quite metal poor compared to giant
ellipticals, which accounts for its bluer SED. Actually, the
computation of the theoretical Mg$_2$ index for a composite stellar
population with the Bulge MD as determined here yields values as low
as $0.16-0.19$ for ages ranging from 10 to 15 Gyr (Maraston et al.
2002). These values, appropriate for simple stellar
population with [Fe/H]$\simeq -0.5$, result from the effect of the
well populated subsolar metallicity component on the MD (Greggio
1997). At optical wavelenghts, this component provides more relative
flux than the high $Z$ one, which results in both a lower composite
Mg$_2$ index, and in a bluer SED.

\subsection{The Sampled Luminosity-Star Number Connection}

The number of stars with mass in the range 0.15 to 1 M$_\odot$
in the observed SOFI field can be obtained by
integrating the bulge IMF with the appropriate scale factor A:
\begin{equation}
N_{\rm PRED}=A\int_{0.15}^1 M^{-1.33}dM,
\end{equation}
with $A\propto L_{\rm T}$ (Renzini 1998), where L$_{\rm T}$ is the
bolometric luminosity sampled by the SOFI field. For given age, the
ratio $A/L_{\rm T}$ is a weak function of both metallicity and IMF.
Its dependence on age is relatively stronger, in fact,
from the models by Maraston (2002), at [M/H]=$-0.2$ and for IMF slope
of $-1.33$, $A/L_{\rm T}$ ranges from 0.82 to 1.1 for an age varying
from 10 to 13 Gyr.

The total bolometric luminosity sampled by the SOFI-LARGE field can be
obtained applying the appropriate bolometric corrections to individual
stars in the sample. To evaluate that, we have run the simulation code
to obtain the bolometric and monocromatic luminosities of a stellar
population with the observed metallicity distribution, obtaining:
$L_{\rm T}=1.39 L_V=1.16 L_I=0.70 L_J=0.53 L_H=0.51 L_K$ with the
coefficients changing by at most $3\%$ when varying the age from 10
to 13~Gyr.  No observational errors were applied in this run.

Using the five monochromatic luminosities with the above relation and
averaging the results we obtain a total luminosity of $L_{\rm
T}=177,000 L_\odot$, hence $N_{\rm PRED}=380,000$ and $510,000$,
respectively for 10 and 13~Gyr. These numbers compare with the
424,000$\pm 81,000$ stars in the SOFI-LARGE field, the error in the
latter being dominated by the Poisson noise in the number of objects
in the NICMOS field. Predicted and observed numbers are quite
consistent (see also Paper~I), since these theoretical estimates are
expected to be accurate to within $\sim 10\%$.

\subsection{The $M/L$ Ratios}

Integrating the IMF from 0.15 to 100 $M_\odot$, and adopting the
prescription in Paper~II for the mass of the stellar remnants (those
above $\sim 1M_\odot$) one derives the total stellar mass in the bulge
SOFI-LARGE sample. Using the luminosities in the various bands given
above, one then determines the corresponding $M/L$ ratios. This gives:

\smallskip\par\noindent
$M/L_{\rm V}=3.67$\par\noindent
$M/L_{\rm I}=3.25$\par\noindent
$M/L_{\rm J}=1.28$\par\noindent
$M/L_{\rm H}=1.00$\par\noindent
$M/L_{\rm K}=0.87$,\par\noindent
\smallskip\par\noindent
with $M/L_{\rm K}$ in very nice agreement with the dynamical value,
$M/L_{\rm K}\simeq 1$ (Kent 1992).

\section{The Age of the Bulge}

Taking advantage of the sharp turnoff region of the decontaminated
near-IR CMD derived in Section~3 we proceed to estimate the age of the
bulge stellar population. As in paper I, we adopt a differential
procedure, comparing the luminosity difference between the HB clump and the
MS turnoff of the bulge to that of a globular cluster of similar metallicity.

\begin{figure}
\centerline{\psfig{figure=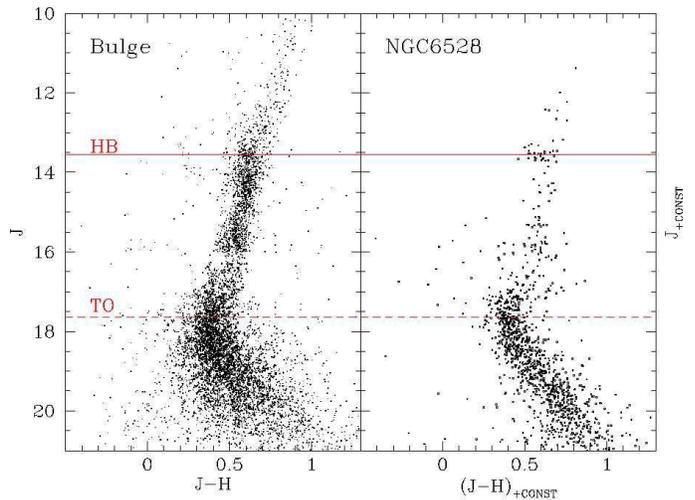,width=9cm,angle=-90}}
\caption{The bulge and the NGC 6528 CMDs are shown side by side.
The magnitude difference between the HB and the turnoff of in the bulge
CMD (left) is compared with the same quantity for NGC~6528
(right). The CMD for NGC~6528, originally in the NICMOS instrumental
(m$_{110}$,m$_{110}$-m$_{160}$) plane, has been shifted both in
magnitude and in color in order to match the location of the bulge
HB.}
\label{age}
\end{figure}

Fig.~\ref{age} shows the comparison between the bulge CMD and that of
the cluster NGC~6528, whose metallicity is close to the average of the
bulge.  The near-IR CMD of NGC~6528 is based on the NICMOS photometry
obtained by Ortolani et al. (2001). The magnitude difference between
the HB clump and the turnoff is virtually identical in the two
diagrams, as emphasized by the two horizontal lines. 

This ensures that the difference between the age of the cluster and
the {\it mean} age of the bulge cannot exceed $\sim 20\%$ (thanks to
the {\it rule of thumb} according to which $\delta$age/age$\simeq
\delta(\Delta M^{\rm HB}_{\rm TO})$ (Renzini 1991).

This confirms and reinforces the conclusion in Paper~I that the bulk
of the bulge population and the clusters NGC~6528 and NGC~6553 are
coeval.  The absence of any appreciable extension of the bulge main
sequence beyond the obvious turnoff makes it clear that no trace of an
intermediate-age population is detectable in the bulge CMD.

\begin{figure}
\centerline{\psfig{figure=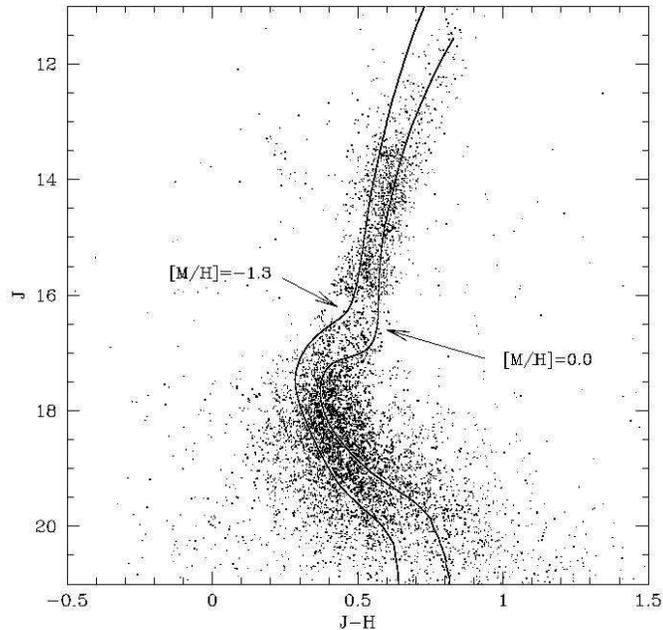,width=9cm}}
\caption{10~Gyr isochrones (Cassisi \& Salaris 1997) for the two extremes 
of the bulge MD are overplotted on the CMD.}
\label{ageiso}
\end{figure}

The proper motion decontaminated and differential reddening corrected
CMD of NGC~6553 (Zoccali et al. 2001a) and NGC 6528 (Feltzing et
al. 2002) confirm the results of Paper~I that the HB to TO magnitude
difference in these two bulge clusters is virtually identical to that
of the inner halo clusters NGC~104 (47~Tuc). Fig.~\ref{age} now shows
that this magnitude difference is essentially identical also for the
bulge field population, strengthening the case for the bulk of the
whole population of the Galactic spheroid (i.e. bulge and halo) being
essentially coeval, though an age difference of $\sim 20\%$ ($\sim
2-3$ Gyr) either way cannot be excluded.

One aspect of the cluster to bulge CMD comparison still deserves some
attention. Indeed, the bulge population is affected by dispersion in
both distance and metallicity, while the cluster stars are chemically
homogeneous and all at the same distance (although affected by some
differential reddening). In Fig.~\ref{ageiso} two 10~Gyr isochrones
spanning the full metallicity range of the bulge are overplotted to
the bulge CMD, assuming the same distance and reddening for both of
them.  This illustrates that the wider dispersion affecting the bulge
CMD (compared to the HST/NICMOS CMD of NGC~6528) can be well accounted
by the bulge metallicity dispersion, also taking into account the
$\sim 0.13$ mag 1-$\sigma$ dispersion due to the distance distribution
along the line of sight.

According to recent attempts at determining the relative ages of
Galactic globular clusters the bulk of clusters are coeval within a
$\pm 1.5$ Gyr uncertainty, with only the most metal rich ones in the
sample appearing to be slightly younger than the others (Rosenberg et
al. 1999; Salaris \& Weiss 2002). However, these studies do not extend to
the high-metallicity clusters of the bulge. For example, Rosenberg et
al. assign to 47~Tuc an age $1.2\pm1.2$ Gyr ``younger'' than that the
bulk of the halo globular clusters. Salaris \& Weiss (2002) assign to
the same cluster an age of $10.7\pm 1.0$ Gyr, compared to $11.7\pm
0.8$ Gyr for the prototypical metal poor cluster NGC 7078 (M15). On
the other hand, Ortolani et al. (2001) date NGC 6528 at 13$\pm3$ Gyr
from the value of $\Delta J^{\rm HB}_{\rm TO}$.

\begin{figure}
\centerline{\psfig{figure=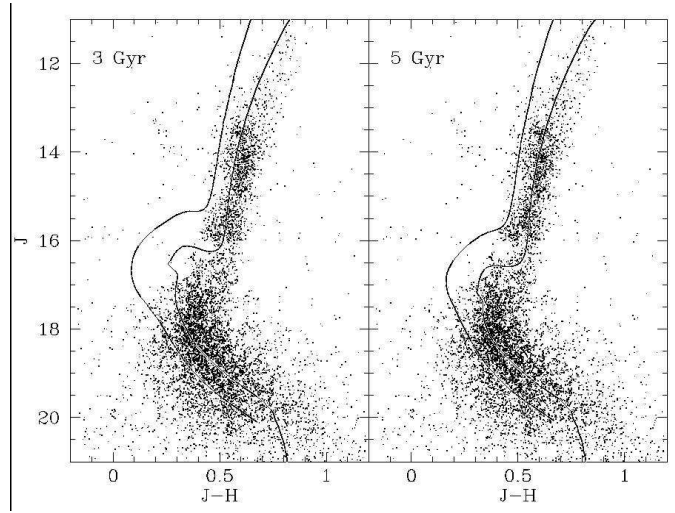,width=9cm,angle=-90}}
\caption{Comparison of the bulge CMD with two younger isochrones of
3 (left panel) and 5 Gyr (right panel). Two models are plotted in
each panel, both referring to the same age. The reddest curve in
each panel is for solar metallicity, while the one on the blue 
side is for [M/H]=$-1.3$.}
\label{age_young}
\end{figure}

It is clear that, within the uncertainties of the currently available
data and dating methods, no appreciable age difference has been
unambiguously detected between the bulk of bulge field stars and the
globular clusters of either the bulge or the halo.  On the other hand,
the {\it absolute} age of the clusters remain more uncertain than the
formal error bars sometime quoted by individual authors.  Just to
mention one example, the age of the globular cluster 47~Tuc has been
recently estimated to be $12.5\pm2$ (Carretta et al. 2000),
$13\pm2.5$ Gyr (Zoccali et al. 2001b), and $10.7\pm1.0$ Gyr (Salaris
\& Weiss 2002), the difference being partly due to a difference in the
cluster distance and partly to the use of different sets of models.

\begin{figure}
\centerline{\psfig{figure=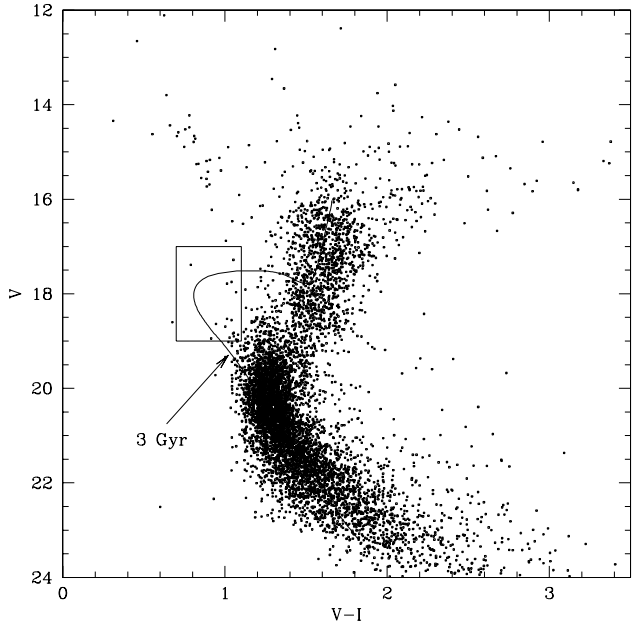,width=9cm}}
\caption{The disk decontaminated optical CMD, together with the 3 Gyr
isochrone for solar metallicity. The box shows the region where
the stars of a hypotetical 3 Gyr old ``bar'' population, spread 
along the line of sight, would be expected.}
\label{cmdvi}
\end{figure}

Significantly younger ages can be excluded, as shown in
Fig.~\ref{age_young}, where 3 and 5 Gyr isochrones of both solar and
[M/H]=$-1.3$ metallicity are overplotted on the bulge CMD.  After the
submission of this paper, during the refereing process, we became
aware of the paper by Cole \& Weinberg (2002), in which the Authors
argue that the bulk of the stellar population of the Galactic ``bar''
formed less than 6 Gyr ago, with an age of $\sim 3$ Gyr being
favored. As they state ``the main sequence turnoff of a 3 Gyr old
population should be readily traceable along the Galactic bar from
$V\approx17$ at the near end to $V\approx19$ at the far end''. Note
that the Galactic component called ``bar'' in Cole \& Weinberg (2002)
has a mass of $2\times 10^{10} M_\odot$ and therefore is not a minor
component, but rather the whole population of the system called here
``the bulge''.  As evident from Fig.~\ref{cmdvi}, no such intermediate
age population is actually detected in the present data.

The region in the CMD just above the main sequence turnoff is so
devoid of stars that very few, if any, blue stragglers stars (BSS) may
be present in the field (see, e.g., Fig.\ref{ageiso}). Among Galactic
globular clusters, Ferraro et al. (1995) estimate an average frequency
of $\sim 1$ BSS every $10^3 L_\odot$ of bolometric light of the parent
cluster, but with very large cluster to cluster variations that are
not merely statistical fluctuations.  Scaling from the SOFI-LARGE
field, the SOFI-SMALL field samples $\sim 177,000/4.6\simeq 38,000\;
L_\odot$, and one would expect to recover $\sim 38$ BSSs, if the bulge
has the same BSS frequency as the average globular cluster. Clearly it
has not. The bulge is far less productive of BSSs than a typical
globular cluster, indicative that the cluster environment favors the
formation of binaries with the right separation for producing
BSSs. Most likely this is due to the dynamical processes that are
germaine to the clusters.


\section{The RGB Tip and Above}

The tip of the RGB is of special interest for three main reasons: i)
any AGB star brighter than the RGB tip is a candidate intermediate age
star, hence may signal the presence of an intermediate age population
(e.g. Iben \& Renzini 1983), ii) the RGB tip itself is directly used
as a distance indicator (Lee, Freedman \& Madore 1993; Salaris \&
Cassisi 1997; Sakai, Madore, \& Freedman 1999), and iii) in any galaxy
the luminosity function near the tip of the RGB has a direct impact on
the distance determination using the surface brightness fluctuation
method (Tonry \& Schneider 1988).

\begin{figure}
\centerline{\psfig{figure=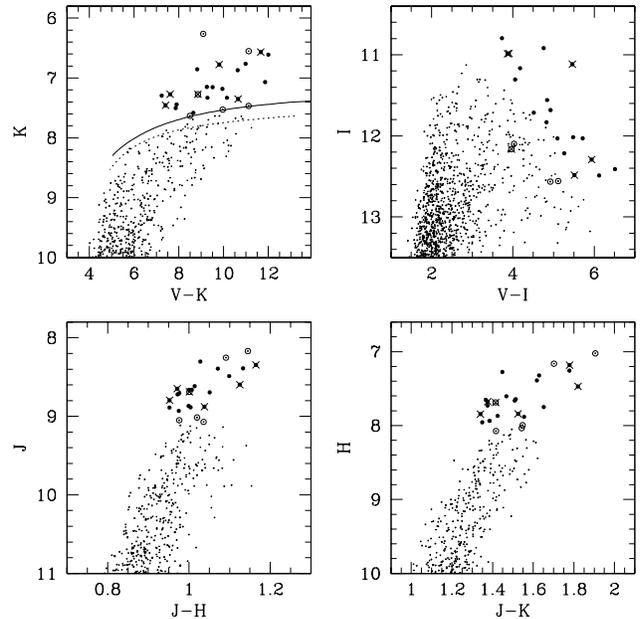,width=9cm}}
\caption{The bulge RGB tip in different bands. The solid line is the
theoretical tip location according to Salaris \& Cassisi (1998),
while the dashed line is the location of the tip empirically found
by Ferraro et al. (2000). Big dots are the (20) stars that lie above
the theoretical RGB tip in {\it all} the bands, while open circles
are stars that lie above the tip in $K$ but not in $I$.
Crosses are blends even in the optical images.}
\label{RGBtip}
\end{figure}

Theoretical models indicate that the bolometric luminosity of the RGB
tip (corresponding to the helium flash in the core) increases with
metallicity by $\sim 0.25$ mag per dex (Rood 1972; Sweigart \& Gross
1978). However, due to increasing blanketing, in the $I$ band the RGB
tip magnitude appears to be independent of metallicity for
[Fe/H]$\lsim -0.7$, i.e., in the range covered by
globular clusters in the Galactic halo (Lee et al. 1993). The present
data on the Galactic bulge offer an opportunity to explore the
behavior of the RGB tip luminosity at higher metallicities, up to
solar and beyond.

Fig.~\ref{RGBtip}a displays the $(K,V-K)$ diagram of the combined
WFI+2MASS field, for the upper RGB+AGB ($K<10$). Overplotted, as a
solid line, is the location of the theoretical RGB tip for
metallicities in the range from [M/H]=$-2.35$ to [M/H]=$-0.28$
(extrapolated here to higher metallicity).  For each metallicity, the
RGB tip bolometric luminosities from the theoretical models of Salaris
\& Cassisi (1998) have been converted to $K$ magnitudes and $V-K$
colors using the empirical bolometric corrections from Montegriffo et
al.  (1998) and the analytical template RGBs described in
Section~4. The empirical location of the RGB tip in the template
globular clusters is also shown as a dotted line (from Ferraro et
al. 2000), with the small offset between the theoretical and empirical
RGB tips being probably due to poor statistics near the RGB tip in the
template globular clusters.

In Fig.~\ref{RGBtip}, 25 stars lie above the nominal RGB tip in $K$,
and therefore are candidate AGB stars. Three more stars with very red
colors ($J-K>2.5$) lie outside the limits of these plots.  Two of them
are included in the IRAS catalogue (IRAS18061-3140 and IRAS18069-3153)
as OH-IR stars. The third one, missing in our optical CMD due to
incompleteness ($V$ mag for the other two stars is 20.7 and 23.7,
respectively) is likely to be of the same nature. Two stars in the
same CMD region are also present in the disk CMD from 2MASS, therefore
suggesting that they are not tracers of a younger bulge population.
As well known, in metal poor globular clusters ([Fe/H]$\lsim -1$) no
AGB stars brighter than RGB tip have been found. However, a few stars
brighter than the RGB tip do exists in more metal rich globulars, with
many of them being long period variables (LPV, Frogel \& Elias 1988;
Guarnieri et al. 1997). The frequency of LPVs in metal rich globulars
has been evaluated by Frogel \& Whitelock (1998). From their Table 1,
one derives an LPV frequency of $\sim 4/(10^6L_{\rm K}^\odot$). The
luminosity sampled by the combined WFI-2MASS field is 13.43 times
larger than that sampled by the SOFI-LARGE field, i.e.  $\sim
5.7\times 10^6 L_{\rm K}^\odot$, hence one would expect 23 LPVs, while
28 stars are found above the RGB tip.  However, an inspection of their
optical counterparts in the WFI frames reveals that 6 of them, shown
as crosses in Fig.~\ref{RGBtip}, are blended with other dimmer stars,
and their intrinsic $K$-band luminosity will be somewhat dimmer than
estimated on 2MASS data. Another 5 stars are definitely below the RGB
tip in the optical CMD (Fig.~\ref{RGBtip}b), and are plotted as open
circles in the four panels of Fig.~\ref{RGBtip}. Their location may
result either from the resolution of 2MASS being much worse ($\sim 4$
arcsec) than that of WFI ($\lsim 1$ arcsec) and therefore these stars
may have unresolved companions in near-IR, that make them artificially
brighter.  Alternatively they might be LPVs having been caught near
minimum light at the time of WFI observations. Finally, our
simulations show that $\sim 5$ RGB stars are expected to be found
above the nominal RGB tip due to the depth effect plus differential
reddening. On the other hand, other LPVs may have been caught below
the RGB, near minimum light. All in all we conclude that the number of
stars brighter than the RGB tip is within the expectations as derived
from the frequency of LPVs in globular clusters spanning nearly the
same metallicity range of the bulge. Of course, not all stars brighter
than the RGB tip in Fig.~\ref{RGBtip} may be LPVs, hence the present
findings do not contradict the result of Frogel \& Whitelock (1998),
who argue for a deficit of LPVs in the bulge field compared to metal
rich globulars. In any event, the bright AGB stars seen in the
WFI-2MASS field do not favor the presence of a stellar population in
the galactic bulge appreciably younger than the old, metal rich
globular clusters, thus reinforcing the conclusion we have drawn from
from the turnoff region.

\begin{figure}
\centerline{\psfig{figure=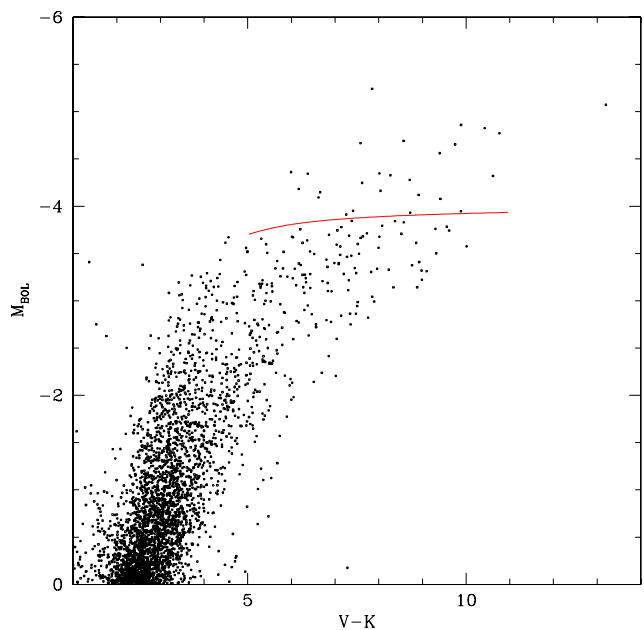,width=9cm}}
\caption{The absolute bolometric luminosity of upper RGB and AGB
stars. The theoretical RGB tip luminosity is also shown as a solid
line (the same as in Fig.~\ref{RGBtip}). The bulge AGB stars appear to
extend $\sim 1$ bolometric magnitude above the RGB tip.}
\label{RGBol}
\end{figure}

The absolute bolometric magnitudes for the stars on the upper RGB and
AGB have also been derived using the empirical bolometric corrections
from Montegriffo et al. (1998) and the distance and reddening adopted
in Section 4.  The result is displayed in Fig.~\ref{RGBol}, showing
that the bulge AGB stars extend $\sim 1$ bolometric magnitude above
the RGB tip, reaching $M_{\rm bol}\simeq -5.0$. This is just
marginally brighter (by $\sim 0.2$ magnitudes) than the brightest
known LPV member of a metal rich globular cluster, i.e. V4 in NGC 6553
(Guarnieri et al. 1997).

Concerning the use of the RGB tip as a standard candle,
Fig.~\ref{RGBtip}b shows that the $I$ band magnitude of the tip drops
by almost two magnitudes, from the lowest to the highest metallicity
spanned by bulge stars. Clearly the $I$ band luminosity of the tip
cannot be used as a standard candle above [Fe/H]$\simeq -0.7$. Also in
the $J$ band the tip luminosity appears to decrease slightly
(Fig.~\ref{RGBtip}c), while it increases slightly in the $K$ band
(Fig.~\ref{RGBtip}a). In the $H$ band instead the tip luminosity
appears to be fairly constant, suggesting the use of this band for
deriving RGB tip distances of stellar populations in the metallicity
range $-1.0\lsim$[Fe/H]$\lsim0.0$. For even higher metallicities the
$K$ band may become more appropriate, as indicated by the theoretical
RGB tip luminosity shown in Fig.~\ref{RGBtip}a.

Fig.~\ref{RGBtip} suggests also quite obvious considerations
concerning the surface brightness fluctuation method. Clearly, the
fluctuation magnitude will depend both on the particular band used,
and on the specific metallicity distribution in each galaxy.

\section{Summary and Conclusions}

Optical and near-IR CMDs for the Galactic bulge
have been presented based on data taken with WFI at the ESO/MPG 2.2m
telescope and with SOFI at the ESO NTT, respectively. This dataset has been
supplemented by the NICMOS photometry from Paper~II, and by public 2MASS
data. The near-IR CMD has been statistically decontaminated from the
disk stars in the foreground, producing a very clean CMD of the bulge.
To ensure great statistical significance also for the upper RGB, the
SOFI data have been combined with 2MASS data on a larger field
including the SOFI field itself, and the 2MASS CMD has also been
statistically decontaminated from the foreground stars.

By combining the best near-IR data in each corresponding luminosity
range (2MASS, SOFI, and NICMOS) a luminosity function of the Galactic
bulge has been constructed that extends over $\sim 15$ magnitudes,
from a few bright AGB stars above the RGB tip, down to almost the
bottom of the main sequence, corresponding to stars of $\sim
0.15\,\msun$. This is the most extended and complete luminosity
function so far obtained for the galactic bulge, hence for galactic
spheroids in general.

Combining WFI optical data and 2MASS near-IR data, a disk
decontaminated $(K,V-K)$ CMD was constructed that includes 503 stars
brighter than $M_{\rm K}=-4.5$.  Together with empirical RGBs of
globular clusters of known metallicity used as templates, this fairly
large sample of bulge stars is used to determine the bulge metallicity
distribution. It is found to peak at [M/H]$\sim -0.1$, with a fairly
sharp edge just above solar metallicity, and a low-metallicity tail
that does not appreciably extend below [M/H]$\sim -1$, quite similar
to the spectroscopic result of McWilliam \& Rich (1994), but relative
to a much larger sample. Such a distribution contains somewhat less
metal poor stars than predicted by the {\it closed box model} of
chemical evolution (e.g., Rich 1990), and indicates that the classical
{\it G-dwarf problem} may affect also the Galactic bulge, as known
since a long time for the disk (e.g., Tinsley 1980). Integrated light
studies indicate that a G-dwarf problem is also present in early type
galaxies, as a closed-box model metallicity distribution will result
in too low values of the Mg$_2$ index (Greggio 1997). Yet, the reasons
for both disks and spheroids sharing the same problem are likely to be
at least in part quite different.  For the Galactic disk (and perhaps
disks in general) pre-enrichment by the older bulge is likely to set a
floor metallicity at [Fe/H]$\sim -1.0$ (Renzini 2002). Clearly, this
kind of pre-enrichment cannot be invoked for the bulge itself.

In the closed-box model one assumes the whole baryonic mass to be
already assembled at $t=0$, while by adopting a gradual built up the
G-dwarf problem can be eliminated. In the case of the Galactic disk
this solution of the problem has taken the name of {\it infall}
(e.g. Pagel 1997, and references therein). Since the star-forming ISM
(either in a disk or in a proto-spheroid) is likely to be built up
gradually, then the closed-box model would not apply.  This situation
is indeed established both in the case of continuous gas infall
forming disks, as well as in the case of early, merging-driven
starbursts building galactic spheroids. In turn, starbursts are likely
to drive metal-rich galactic winds, thus discontinuing star formation
before the gas is fully turned into stars. Therefore, the bulge likely
departed from the closed-box approximation both for having being built
gradually, albeit rapidly, and for having ejected gas and metals near the
climax of its star formation activity.

The near-IR CMD of the Galactic bulge, carefully decontaminated from
foreground stars, has been used to show that the bulge is virtually
coeval with the Galactic halo. The
region in the CMD above the main sequence turnoff of the bulge
is so devoid of stars that no trace can be found of a
population of bulge stars significantly younger than the main old
component. This indicates that no appreciable new stars formed in
the bulge, or were added to it, following the intense starburst
activity that turned $\sim 10^{10}\msun$ of gas into the bulge stars,
some 10 Gyr ago.

Using both optical and near-IR data, the RGB tip of the bulge is
clearly identified, and its systematics in the various bands is
established.  Contrary to the behavior at lower metallicities, in the
metallicity range spanned by the bulge stars ($-1\lsim{\rm [Fe/H]}\lsim 0$)
the $I$-band luminosity of the RGB tip is not constant, but decreases
with increasing metallicity.  Conversely, the RGB tip luminosity is
found to be constant in the $H$ band, while it increases with
metallicity in the $K$ band. This behavior is naturally accounted for
by a combination of increasing blanketing which is stronger at short
wavelengths (i.e. in the $I$ and $J$ band), with the bolometric
luminosity of the RGB tip which increases with increasing
metallicity. The metallicity distribution of galactic spheroids is
therefore expected to affect the use of the surface brightness
fluctuation method of distance determinations (Tonry \& Schneider
1988). As a final remark, we would like to note that a group of stars
is found at luminosities exceeding the RGB tip, and their number is
roughly consistent with the expectation scaling from old, metal rich
globular clusters. Their number is consistent with the purely
old age for the bulge population as derived from the turnoff region.

Worth mentioning are also some limitations of the present study.  The
location of the peak of the metallicity distribution coincides with
the metallicity of the clusters NGC 6528 and NGC 6553, but in turn the
metallicity itself of these clusters remain quite
uncertain. Systematic, homogeneous, high spectral resolution studies
of stars in these clusters and in the field are needed to properly
match the cluster and field star metallicity scales. When such data
will become available, it will be easy to re-derive the metallicity
distribution of the bulge. Finally, we regret to have been unable
to find the origin of the larger HB and RGB spread exhibited by the
bulge CMD, compared to simulations that are meant to include all
intrinsic spreads of the real bulge population (distance, metallicity,
and reddening).

In closing, we would like to note that the Milky Way galaxy is a rather
uncospicuous late-type spiral, member
of a poor group of galaxies located quite away from major density
peaks in the distribution of galaxies. Yet, virtually its whole
spheroidal component, from the outer metal-poor halo all the way to
the central metal-rich bulge, is $\sim 1$ Hubble time old. There is
now conclusive evidence that the stellar populations of galactic
spheroids in general, elliptical and bulges, are also $\sim 1$ Hubble
time old (e.g. Renzini 1999, for an extensive review), while most such
spheroids appear to be well in place and fully assembled already at
$z\sim 1$, having passively evolved since at least $z\sim 2.5$
(Cimatti et al. 2002). If the stellar mass in galactic spheroids amounts
to 50-70$\%$ of the total stellar mass in the local universe (Fukugita et al.
1998), all this evidence indicate that a major fraction
of all stars formed at high redshift, in deep potential wells bound to
become the galactic spheroids of today. If such build up of galaxies
was promoted by the hierarchical merging of dark matter halos, it took
place at a time and at a rate that no practical rendition of the
hierarchical merging paradigm has yet succeded to reproduce (Cimatti
et al. 2002 and references therein).

\begin{acknowledgements}

We thank the referee, Dr. E.M. Sadler, for careful reading of the manuscript 
and comments that lead to an improvement of the paper.

The present work has been partially supported by the italian Ministero
della Universit\`a e della Ricerca under the program {\it ``The Origin
and Following Evolution of the Stellar Population in the
Galactic Spheroid''} and by the Agenzia Spaziale Italiana.

S.C. acknowledges the financial support by (Cofin2001) under the
scientific project: ``Origin and evolution of stellar populations in
the galactic spheroid''

RMR acknowledges support from grant number AST-0098739, from the
National Science Foundation and from grant GO-7891 from the Space
Telescope Science Institute.

\end{acknowledgements}



\begin{thebibliography}{}

\bibitem{} Allen's Astrophysical Quantities, 2000, 4th edition, ed. Cox, A.N.
	(New York: Springer-Verlag)
\bibitem{} Barbuy, B. 1999, Ap\&SS, 265, 319
\bibitem{} Barbuy, B., Renzini, A., Ortolani, S., Bica, E., \& Guarnieri, M.D.,
	1999, A\&A 341, 539
\bibitem{} Bono, G., Caputo, F., Cassisi, S., Castellani, V., \& Marconi, M. 1997,
	ApJ, 479, 279
\bibitem{} Cardelli, J.\ A., Clayton, G.\ C., \& Mathis, J.\ S. 1989, ApJ,
        345, 245
\bibitem{} Carney, B.W. 1996, PASP, 108, 900
\bibitem{} Carretta, E., Cohen, J.G., Gratton, R.G., Behr, B.B. 2001, AJ, 122,
           1469
\bibitem{} Carpenter, J.M. 2001, AJ, 121, 2851
\bibitem{} Cassisi, S., Castellani, V., Ciarcelluti, P., Piotto, G., \& Zoccali, M.
	2000, MNRAS, 315, 679
\bibitem{} Cassisi, S., Salaris, M., \& Bono, G. 2002, ApJ, 565, 1231
\bibitem{} Cassisi, S., \& Salaris, M. 1997, MNRAS, 285, 593
\bibitem{} Castelli, F., Gratton, R.G., Kurucz, R.L., 1997, A\&A, 318, 841
\bibitem{} Cimatti, A., Daddi, E., Mignoli, M., Pozzetti, L., Renzini, A.
           et al. 2002, A\&A (in press, astro-ph/0111527)
\bibitem{} Coelho, P., Barbuy, B., Perrin, M.-N., Idiart, T., Schiavon, R.P.,
	Ortolani, S., Bica, E., 2001, A\&A, 376, 136
\bibitem{} Cohen, J.G., Gratton, R.G., Behr, B., Carretta, E., 1999, ApJ, 523, 739
\bibitem{} Cole, A.A., \& Weinberg, M.D. 2002, ApJ, 574, L43
\bibitem{} Cot\'e, P. 1999, AJ, 118, 406
\bibitem{} Davidge, T.J. \& Nieto J.-L., 1992, ApJ, 391, L13
\bibitem{} Depoy, D.L., Terndrup, D.M., Frogel, J.A., Atwood, B., Blum, R. 
	1993, AJ, 105, 2121
\bibitem{} Elston, R., \& Silva, D.R., 1992, AJ, 104, 1360
\bibitem{} Faber, S.M., Friel, E.D., Burstein, D., \& Gaskell, C.M. 1985,
	ApJS, 57, 711
\bibitem{} Feltzing, S. \& Gilmore, G. 2000, A\&A, 355, 949
\bibitem{} Feltzing, S., Johnson, R.A., \& de Cordova, A. 2002, A\&A, 385, 67
\bibitem{} Ferraro, F., Fusi Pecci, F., \& Bellazzini, M. 1995, A\&A, 294, 80
\bibitem{} Ferraro, F.R., Montegriffo, P., Origlia, L., \& Fusi Pecci, F. 2000,
	AJ, 119, 1282
\bibitem{} Forbes, D.A., Brodie, J.P., \& Larsen, S. 2001, ApJ, 556, L83
\bibitem{} Freedman, W.L., 1992, AJ, 104, 1349
\bibitem{} Frogel, J.A., \& Elias, J.H. 1988, ApJ, 324, 823
\bibitem{} Frogel, J.A., \& Whitelock, P.A. 1998, AJ, 116, 754
\bibitem{} Frogel, J.A., \& Whitford, A.E. 1987, ApJ, 320, 199
\bibitem{} Frogel, J.A., Stephens, A., Ramirez, S., DePoy, D.L. 2001, AJ, 122, 1896
\bibitem{} Fugugita, M., Hogan, C.J., \& Peebles, P.J.E. 1998, ApJ, 503, 518
\bibitem{} Greggio, L. 1997, MNRAS, 285, 151
\bibitem{} Greggio, L., Tosi, M., Clampin, M., De Marchi, G., Leitherer, C.,
	Nota, A. \& Sirianni, M. 1998, ApJ, 504, 725
\bibitem{} Guarnieri, M.D., Renzini, A., \& Ortolani, S. 1997, ApJ, 477, L21
\bibitem{} Guarnieri, M.D., Ortolani, S., Montegriffo, P., Renzini, A.,
	Barbuy, B., Bica, E., Moneti, A.: 1998, A\&A, 331, 70
\bibitem{} Harris W.E. 1976, AJ, 81, 1095
\bibitem{} Harris W.E. 1996, AJ, 112, 1487
\bibitem{} Harris W.E. 2001, in Star Clusters, ed. L. Labhardt \& B. Binggeli
           (Berlin: Springer-Verlag), p. 223
\bibitem{} Hartwick, F.D.A. 1976, ApJ, 209, 418
\bibitem{} Holtzman, J.A., Light, R.M., Baum, W.A., Worthey, G., Faber, S.M.,
	et al. 1993, AJ, 106, 1826
\bibitem{} Holtzman, J.A., Watson, A.M., Baum, W.A., Grillmair, C.J.,
        Groth, E.J., et al. 1998, AJ, 115, 1946
\bibitem{} Iben, I. Jr. 1968, ApJ, 154, 581
\bibitem{} Iben, I. Jr., \& Renzini, A. 1983, ARA\&A, 21, 271
\bibitem{} Kent, S.M. 1992, ApJ, 387, 181
\bibitem{} Kinney, A.L., Calzetti, D., Bohlin, R.C., McQuade, K.,
        Sorchi-Bergmann, T., \& Schmitt, H.R. 1996, ApJ, 467, 38
\bibitem{} Landolt, A.U. 1992, AJ, 104, 340
\bibitem{} Lee, M.G., Freedman, W., \& Madore, B.F. 1993, ApJ, 417, 553
\bibitem{} Mannucci, F., Basile, F., Poggianti, B.M., Cimatti, A., Daddi, E.,
           Pozzetti, L., \& Vanzi, L. 2001, MNRAS, 326, 745
\bibitem{} Maraston, C., Greggio, L., Renzini, A., Ortolani, S., Saglia, R.
	   Puzia T.H., Kissler-Patig M., 2002, A\&A, in press, (astro-ph/0209220)
\bibitem{} McNamara, D.H, Madsen, J.B., Barnes, J. \& Ericksen, B.F. 2000,
       PASP,
	112, 202
\bibitem{} McWilliam, A. \& Rich, R.M. 1994, ApJS, 91, 749
\bibitem{} Minniti, D. 1995, AJ, 109, 1663
\bibitem{} Momany, Y., Vandame, B., Zaggia, S., Mignani, R.P., da Costa, L.,
        et al. 2001, A\&A, 379, 436
\bibitem{} Montegriffo, P., Ferraro, F.R., Origlia, L. \& Fusi Pecci, F.
	1998, MNRAS, 297, 872
\bibitem{} Mould, J.R. 1984, PASP, 96, 773
\bibitem{} Ng, Y.K., \& Bertelli, G. 1996, A\&A, 315, 116
\bibitem{} Oke, J.B., \& Gunn, J.E. 1983, ApJ, 266, 713
\bibitem{} Origlia, L., Rich, R.M., Castro, S. 2002, AJ, 123, 1559
\bibitem{} Ortolani, S., Renzini, A., Gilmozzi, R., Marconi, G., Barbuy,
	E., Bica, E. \& Rich, R.M. 1995, Nature, 377, 701 (Paper~I)
\bibitem{} Ortolani, S., Barbuy, B., Bica, E., Renzini, A., Zoccali, M. et al.
	2001, A\&A, 376, 878
\bibitem{} Pasquini, L., Avila, G., Allaert, W., Ballester, P., Biereichel, P.,
	et al. 2000, SPIE, 4008, 129
\bibitem{} Pagel, B.E.J. 1997, Nucleosynthesis and Chemical Evolution of
         Galaxies (Cambridge University Press), p. 237
\bibitem{} Persic, M. \& Salucci, P. 1992, MNRAS, 258, 14
\bibitem{} Persson, S.E., Murphy, D.C., Krzeminski, W., Roth, M., Rieke, M.J,
	1998, AJ, 116, 2475
\bibitem{} Puzia, T.H., Saglia, R.P., Kissler-Patig, M., Maraston, C., Greggio, L., 
	   Renzini, A., Ortolani, S. 2002, A\&A in press (astro-ph/0209238)
\bibitem{} Ramirez, S.V., Stephens, A.W., Frogel, J.A., DePoy, D.L. 2000, AJ,
	120, 833
\bibitem{} Rejkuba, M., 2002, PhD Thesis (Pontificia Universidad Cat\'olica de Chile)
\bibitem{} Renzini, A. 1991, in Observational Tests of Cosmological Inflation,
           ed. T. Shanks, A.J. Banday, \& R.S. Ellis (Dordrecht: Kluwer),
           p. 131
\bibitem{} Renzini, A. 1997, ApJ, 488, 35
\bibitem{} Renzini, A. 1998, AJ, 115, 2459
\bibitem{} Renzini, A. 1999, in The Formation of galactic Bulges, ed. C.M.
           Carollo, H.C. Ferguson, \& R.F.G. Wyse (Cambridge University
           Press), p. 9
\bibitem{} Renzini, A. 2002, in Chemical Enrichment of Intracluster and
           Intergalactic Medium", ed. F. Matteucci and R. Fusco-Femiano
           ASP Conf. Ser., 253, 331
\bibitem{} Renzini, A. \& Buzzoni, A. 1986
	in Spectral evolution of galaxies, ed. C. Chiosi \& A. Renzini
	(Dordrecht: Reidel), p. 135
\bibitem{} Rich, R.M. 1988, AJ, 95, 828
\bibitem{} Rich, R.M. 1990, in Bulges of Galaxies, ed. B.J. Jarvis \& D.M.
           Terndrup (Garching: ESO), p. 65
\bibitem{} Rich, R.\ M., Ortolani, S., Bica, E., \& Barbuy, B. 1998, AJ, 116, 1295
\bibitem{} Rich, R.M., Origlia, L., \& Castro, S.M.  2001, in preparation
\bibitem{} Rood, R.T., 1972, ApJ, 177, 681
\bibitem{} Rosenberg, A., Saviane, I., Piotto, G. \& Aparicio, A. 1999, AJ,
	118, 2306
\bibitem{} Sadler, E.M., Rich, R.M., Terndrup, D.M. 1996, AJ, 112, 171
\bibitem{} Sakai, S., Madore, B.F., \& Freedman, W. 1999, ApJ, 511, 671
\bibitem{} Salaris, M. \& Cassisi, S. 1996, A\&A, 305, 858
\bibitem{} Salaris, M. \& Cassisi, S. 1997, MNRAS, 289, 406
\bibitem{} Salaris, M. \& Cassisi, S. 1998, MNRAS, 298, 166
\bibitem{} Salaris, M., Cassisi, S. \& Weiss, A. 2002, PASP, 114, 375
\bibitem{} Salaris, M. \& Weiss, A. 2002, A\&A, 388, 492
\bibitem{} Saviane, I., Rosenberg, A., Piotto, G., \& Aparicio, A. 2000, A\&A, 355,
	966
\bibitem{} Searle, L. \& Sargent, W.L.W. 1972, ApJ, 173, 25
\bibitem{} Sellwood, J, \& Sanders, R., MNRAS, 233, 611
\bibitem{} Steidel, C.C., Giavalisco, M., Dickinson, M. \& Adelberger, K.L. 1996,
	AJ, 112, 352
\bibitem{} Stephens, A.W., Frogel, J.A., Ortolani, S., Davies, R.,
	Jablonka, P., Renzini, A., \& Rich R.M. 2000, AJ 119, 419
\bibitem{} Stetson, P.B. 1987, PASP, 99, 191
\bibitem{} Stetson, P.B. 1994, PASP, 106, 250
\bibitem{} Stetson, P.B. 2000, PASP, 112, 925
\bibitem{} Sweigart, A.V., \& Gross, P.G.1978, ApJS, 36, 405
\bibitem{} Tantalo, R., Chiosi, C., Bressan, A. \& Fagotto, F. 1996, A\&A, 311, 361
\bibitem{} Terndrup, D.M., 1988, AJ, 96, 884
\bibitem{} Tiede, G.P., Frogel, J.A., \& Terndrup, D.M. 1995, AJ, 110, 27 88
\bibitem{} Tinsley, B.M. 1980, Fund. Cosmic Phys. 5, 287
\bibitem{} Tonry, J. \& Schneider, D.P. 1988, AJ, 96, 807
\bibitem{} van Dyk, S. 2000, BeSN, 34, 40
\bibitem{} Zinn, R. 1996, ASP Conf. Ser. 92, 211
\bibitem{} Zoccali, M., Cassisi, S., Bono, G., Piotto, G., Rich, R.M., \&
	   Djorgovski, S.G. 2000, ApJ, 538, 289
\bibitem{} Zoccali, M., Cassisi, S., Frogel, J.A., Gould, A., Ortolani, S.,
           Renzini, A., Rich, R.M., Stephens, A.W. 2000, ApJ, 530, 418
          (Paper~II)
\bibitem{} Zoccali, M., Renzini, A., Ortolani, Bica, E., \& Barbuy, B. 2001a,
           AJ, 121, 2638
\bibitem{} Zoccali, M., Renzini, A., Ortolani, S., Bragaglia, A., Bohlin. R.,
           Carretta, E., Ferraro, F.R., Gilmozzi, R., Holberg, J.B., Marconi,
           G., Rich, R.M., \& Wesemael, F. 2001b, ApJ, 553, 733
\end{thebibliography}
\end{document}